\pdfoutput=1
\documentclass[reprint,twocolumn,superscriptaddress,amsmath,amssymb,amsfonts,aps,prb,floatfix,longbibliography]{revtex4-2}
\usepackage{times}
\usepackage{booktabs}
\usepackage{graphicx}
\usepackage{dcolumn}
\usepackage{float}
\usepackage[makeroom]{cancel}
\usepackage{bm}
\usepackage[colorlinks=true, citecolor=blue, linkcolor=blue, urlcolor=blue]{hyperref}
\usepackage{bbold}
\DeclareMathAlphabet\mathbfcal{OMS}{cmsy}{b}{n}
\usepackage{wasysym}
\usepackage{amssymb}
\usepackage{amsmath}
\usepackage{siunitx}
\usepackage{comment}
\usepackage{array}
\usepackage{multirow}
\usepackage{physics}

\usepackage[usenames,dvipsnames]{xcolor}

\newcommand{\bk}{\mathbf{k}}
\newcommand{\bR}{\mathbf{R}}
\newcommand{\bG}{\mathbf{G}}
\newcommand{\bK}{\mathbf{K}}

\begin{document}

\title{Valley-Helical Superconductivity Driven by Repulsion and Quantum Geometry:\\ Applications to Time-Reversal Symmetric Rhombohedral Graphene}

\author{Julian May-Mann}
\email{maymann@stanford.edu}
\affiliation{Department of Physics, Stanford University, Stanford, CA 94305, USA}
\author{Trithep Devakul}
\affiliation{Department of Physics, Stanford University, Stanford, CA 94305, USA}
 \author{Gal Shavit}
 \affiliation{Department of Physics and Institute for Quantum Information and Matter, California Institute of Technology,
 Pasadena, California 91125, USA}
 \affiliation{Walter Burke Institute of Theoretical Physics, California Institute of Technology, Pasadena, California 91125, USA}

\begin{abstract}
Motivated by the recent experimental signatures of chiral superconductivity in valley-polarized graphene- and transition-metal-dichalcogenide-based systems, we investigate the possibility of chiral pairing in the absence of valley polarization or time-reversal symmetry breaking. For systems with opposite Berry curvature in the two valleys, we find that overscreened repulsion (i.e., the Kohn-Luttinger mechanism) can indeed induce pairing of opposite chiralities in each valley, producing a valley-helical state. Moreover, the valley-helical state can outcompete more conventional intervalley paired states at intermediate coupling, despite only the latter having a weak-coupling instability.
Using a realistic model of time-reversal-symmetric rhombohedral graphene, we find extended ranges of fillings and displacement fields with dominant valley-helical superconductivity. This occurs for all layer numbers considered here, $4\leq N\leq 8$. We show that the valley-helical states can display topologically projected edge modes, unusual magnetic responses, and intertwined Kekulé-like bond-order; features that can be used to distinguish them from more conventional superconductors.
\end{abstract}
\maketitle

{\bf \textit{Introduction.--}}
In superconductors with a time-reversal (TR) invariant normal state, it is typical to consider pairing between TR-related quasiparticle states. The degeneracy of the TR-related states ensures that the paired Fermi-surface(s) are particle-particle nested, and that there is always a weak-coupling instability for superconductivity (SC)~\cite{bennemann2008superconductivity, schrieffer2018theory, shankar1994renormalization}.  In systems with TR-related valleys (e.g., graphene-based systems),~\cite{min2008electronic, castro2009electronic, fuhrer2010graphene}, the above intuition suggests dominant intervalley SC  (i.e., pairing between opposite valleys) when TR is intact.
Conversely, there is no generic weak-coupling instability for intravalley SC (i.e., pairing within the same valley). The Fermi-surface(s) of a given valley are generally not particle-particle nested due to, e.g., trigonal warping ~\cite{kechedzhi2007influence,zhang2010band}, and interactions must therefore exceed a threshold to produce intravalley SC~\cite{chou2025intravalley, sboychakov2025superconductivity, gaggioli2025spontaneous}.

Despite the lack of a weak-coupling instability, recent experiments have found evidence of intravalley SC arising from a TR-broken normal state in two-dimensional van der Waals (vdW) materials; rhombohedral stacked $N$-layer graphene (R$N$G)~\cite{LongJu_pentaSC_Han2025, dutta2026reconfigurable, sheekey2026visualizing, kalantre2026fermiology, hua2026multi} and twisted bilayers of MoTe$_2$~\cite{tmote2_SC_xu2026signaturesunconventionalsuperconductivitynear}. In both cases, the normal state appears to be a spin-valley polarized band with finite Berry curvature. The observation of intravalley SC from TR-broken normal states raises an immediate question: can intravalley SC also arise from a TR-symmetric normal state,
overwhelming the intervalley SC instability?

In this work, we answer this question in the affirmative. For TR-symmetric systems with valley-contrasting Berry curvature and a Kohn-Luttinger-like pairing mechanism---where Coulomb repulsion is overscreened and becomes attractive~\cite{kohn1965new, ghazaryan2021unconventional, yang2025topological, chen2025intrinsic}---we find that intravalley pairing channels outcompete the intervalley channels at intermediate coupling strength.
The resulting intravalley SC has a \textit{valley-helical} structure, where each valley hosts a chiral superconducting order parameter with opposite angular momentum~\cite{kallin2016chiral}. A perpendicular magnetic field can couple to this angular momentum, producing unusual magnetic responses~\cite{sigrist1991phenomenological}. The valley-helical state also induces a Kekulé-like bond-density-wave as a secondary order~\cite{chamon2000solitons, metlitski2010instabilities, fradkin2015colloquium, wang2026kekule}. Additionally, the valley-helical SC can display TR-protected topology, hosting a protected pair of counterpropagating edge modes\cite{tanaka2011symmetry,ZHANG_Kane_Mele_Tritops_PhysRevLett.111.056402,sato2017topological,TRITOPS_HAIM20191}.

\begin{figure}
    \centering
    \includegraphics[width=.9\linewidth]{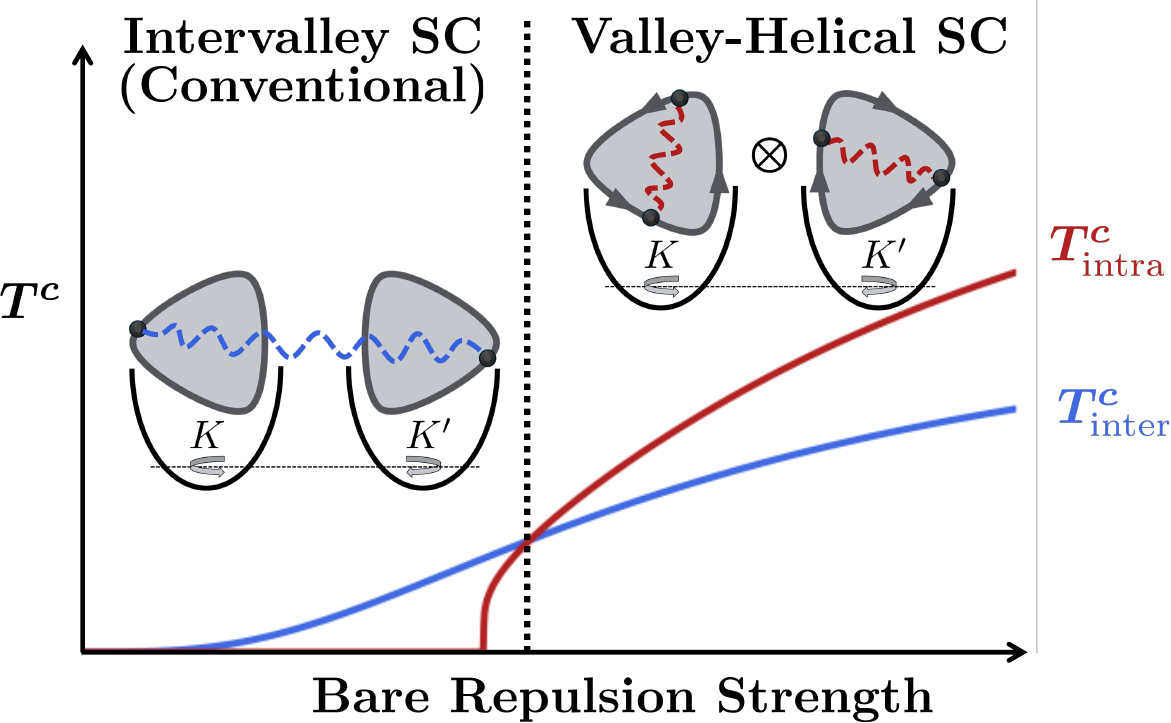}
    \caption{\textbf{Competition between intravalley and intervalley pairing channels.}
    Schematic plot of the intravalley (red) and intervalley (blue) transition temperatures for a two-valley system with valley-contrasting Berry curvature and a Kohn-Luttinger-like pairing mechanism. Since the Fermi surfaces in each valley are generically not particle-particle nested, the intravalley superconductivity requires the repulsion to exceed a threshold, which is set by the strength of trigonal warping and related normal state properties.
    The intravalley channel quickly becomes dominant upon exceeding this threshold.
    }
    \label{fig:schematicTcvsV}
\end{figure}

We show that valley-helical SC arises in these systems because the overscreened pairing interaction remains strongly repulsive at short distances. This short-range repulsion penalizes the intervalley channels in a process we refer to as \textit{residual repulsion poisoning}.
However, certain intravalley channels can avoid the poisoning, as the finite Berry curvature of the paired quasiparticles leads to Cooper pairs with larger inter-particle separation~\cite{maymann2025pairingmechanismdictatestopology}.

For realistically modeled R$NG$ with a symmetric (full-metal) normal state, we find broad ranges of densities and displacement fields with dominant valley-helical SC. Our findings are qualitatively insensitive to the number of layers considered, though quantitative differences do arise. In the main text, we will focus on 7-layer rhombohedral graphene (R$7$G), which appears to be a particularly promising candidate for realizing this phase. When the trigonal warping is large, it cuts off the intravalley instability, allowing intervalley SC to emerge. From this, we conclude that Kohn-Luttinger-like pairing mechanisms favor valley-helical SC in systems with valley-contrasting Berry curvature, provided that the bare repulsion exceeds a threshold set by trigonal warping and related effects. This is shown schematically in Fig.~\ref{fig:schematicTcvsV}.

{\bf \textit{Intravalley and intervalley gap equations.--}}
For our analysis, we consider a normal state consisting of TR-related bands from opposite valleys with opposite Berry curvature. The many-body Hamiltonian projected onto these bands is
\begin{equation}
    H = \sum_{\sigma\tau{\bf k}} \xi^\tau_{\bf k} \phantom{|}\psi^\dagger_{\sigma \tau{\bf k}}\psi^{\phantom{\dagger}}_{\sigma \tau{\bf k}}
    +
    \frac{1}{2A}\sum_{\bf q} V_{\bf q}\tilde{\rho}_{\bf q}\tilde{\rho}_{\bf -q}, \label{eq:projectedBareHamiltonian}
\end{equation}
where we have assumed that each valley is spin degenerate and SU$(2)$ symmetric (treatment of spin-polarized valleys is analogous, and will be discussed towards the end of the work). $\psi_{\sigma \tau{\bf k}}$ is the band-projected quasiparticle operator of spin $\sigma = \uparrow,\downarrow$, valley $\tau = K,K' \equiv \pm $, and momentum $\bf k$ relative to the valley center. For our present analysis, we consider a realistic dual-gate-screened Coulomb interaction $V_{\bf q}= \frac{e^2}{\epsilon_0 \epsilon_\perp |\bf q|}\tanh(d|\bf q|)$, where $\epsilon_\perp$ is the perpendicular dielectric of the device, and $d $ is the distance to the screening gates. $A$ is the area of the system. Due to TR-symmetry, the dispersions satisfy $\xi^\tau_{\bf k} = \xi^{-\tau}_{-\bf k}$. The projected density operator $\tilde{\rho}_{\bf q}$ is
\begin{equation}
    \tilde{\rho}_{\bf q}=\sum_{\sigma \tau,{\bf k}} \Lambda^\tau_{\bf k,k+q}\psi^\dagger_{\sigma \tau{\bf k}}\psi^{\phantom{\dagger}}_{\sigma \tau{\mathbf{k}+\mathbf{q}}}\,,
    \,\,\,\,\,
    \Lambda^\tau_{\bf k,k'}= \left\langle {\mathsf u}_{\bf k}^\tau|{\mathsf u}_{\bf k'}^\tau\right\rangle.
    \label{eq:projecteddensity}
\end{equation}
where $\left|{\mathsf u}_{\bf k'}^\tau\right\rangle$ is the Bloch-wavefunction of the $\tau$-valley band, and $ \Lambda^{\tau}_{\bf k,k'} = \left(\Lambda^{-\tau}_{\bf -k,-k'}\right)^*$ are the form factors, which encode the Berry curvature and quantum geometry of the system. Details of the Bloch gauge fixing are given in the Supplemental Material (SM)~\cite{supp}. Without loss of generality, we will assume that the $K$-valley has positive Berry curvature. Unnecessary spin labels have been suppressed here and throughout.

We will consider pairing between opposite spins in both the intravalley and intervalley channels. Since each valley is spin-degenerate, we can have either spin-singlet or spin-triplet pairing in both the intravalley and intervalley channels~\footnote{We implicitly assume $S^z = 0$ triplet pairing throughout, although degeneracy exists due to the SU$(2)$ symmetry of each valley.}.
The superconducting instabilities
are encoded in the
linearized gap equation (LGE), which is accurate near the transition temperature:
\begin{equation}
    \Delta_{{\bf k}}^{\tau \tau'}=-\int\frac{d{\bf k'}}{\left(2\pi\right)^{2}}
    u^{\tau\tau'}_{{\bf kk'}}
    \frac{1- f(\xi^\tau_{{\bf k'}}) - f(\xi^{\tau'}_{{-\bf k'}}) }{\xi^{\tau}_{{\bf k'}}+\xi^{\tau'}_{{-\bf k'}}}
    \Delta_{{\bf k'}}^{\tau \tau'},
\label{eq:linearizedGap}\end{equation}
where $\Delta_{{\bf k}}^{\tau \tau'}$ is the gap function for pairing between valleys $\tau$ and $\tau'$. $u^{\tau\tau'}_{{\bf kk'}}$ is the pairing interaction in the Cooper channel, and $f$ is the Fermi-Dirac distribution.
The highest temperature where Eq.~\eqref{eq:linearizedGap} has a non-trivial solution marks the critical transition temperatures: $T^c_{\mathrm{intra}}$ and $T^c_{\mathrm{inter}}$ for $\tau' = \tau$ and $\tau' = -\tau$ respectively. The largest $T^c$ determines the dominant superconducting channel.

For a Kohn-Luttinger-like pairing mechanism, the pairing interaction can be calculated using the random-phase approximation (RPA),
\begin{equation}
\begin{split}
 &u^{\tau\tau'}_{\bf kk'}=\Lambda_{\bf kk'}^\tau\Lambda_{\bf -k,-k'}^{\tau'} V^{\rm{RPA}}_{\bf k-k'}, \phantom{=} V^{\rm{RPA}}_{\bf q} = \frac{V_{\bf q}}{1+\Pi_{\bf q}  V_{\bf q}},
    \label{eq:pairingInt}
\end{split}
\end{equation}
where $V^{\rm{RPA}}$ is the screened interaction due to particle-hole fluctuations.
The form factors dressing $V^{\rm{RPA}}$ account for projection onto the active bands. Assuming a large separation between the active and remote bands, the polarizability $\Pi_{\bf q}$, can be calculated using only the active bands.

The intravalley and intervalley gap equations differ in their pairing interactions $u^{\tau\tau'}$, and particle-particle susceptibilities [see Eq.~\eqref{eq:linearizedGap}].
As we demonstrate below, the pairing strength tends to be larger in the intravalley channel, though the particle-particle susceptibility favors intervalley SC.

\textit{\textbf{Intravalley pairing and residual repulsion poisoning.--}}
The difference between the intervalley and intravalley pairing interactions arises entirely from the form factors that dress $V^{\rm{RPA}}$ in Eq.~\eqref{eq:pairingInt}.
To illustrate the importance of the form factors analytically, we will make two simplifying assumptions: First, that both valleys have a continuous rotation symmetry with respect to the valley origin. Second, that the Bloch wavefunctions in the $K$-valley can be written as holomorphic functions of $\bf{k}$, up to normalization (anti-holomorphic functions for $K'$-valley bands). Holomorphic bands like these are found in many paradigmatic models that feature finite Berry curvature\cite{guinea2006electronic,bandinversion_PhysRevX.14.041040, lee2017band, ledwith2023vortexability, tan2025ideallimitrhombohedralgraphene, soejima2025lambda, bernevig2025berry,han2025exact, desrochers2026electronic, may2026skyrmion}, and have been widely employed to study the universal physics of vdW materials. Under the above assumptions, the form factors take the simple form
\begin{equation}
\Lambda^+_{\bf k k'} = N^{1/2}_{\bf k} N^{1/2}_{\bf k'} \sum^{\ell_{\mathrm{max}}}_{\ell=0} |c_{\ell}|^2 \left( k^* k'\right)^{\ell},
\label{eq:holomorphicFFs}\end{equation}
where $k=k_x+ik_y$, $c_{\ell}$ are complex coefficients, $N^{-1}_{\bf k} = \sum^{\ell_{\mathrm{max}}}_{\ell=0} |c_{\ell}|^2 |\mathbf{k}|^{2\ell}$, and $\ell_{\mathrm{max}}$ denotes the largest power of $k$ that appears in the Bloch wavefunctions. Here and throughout, when discussing features of a single valley, we will focus on the $K$-valley ($\tau = +$), as the $K'$-valley ($\tau = -$) is always related by TR-symmetry.

\begin{figure}
    \centering
    \includegraphics[width=1.0\linewidth]{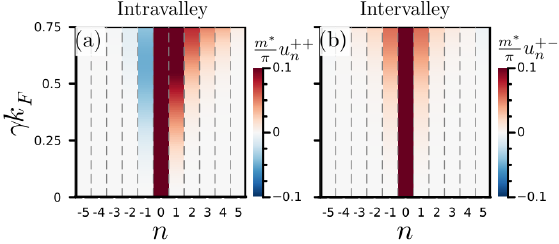}
    \caption{\textbf{Pairing strength in the ideal limit of R$7$G.} (a) Coefficients from the angular momentum decomposition of the intravalley pairing interaction for ideal R$7$G as a function of $\gamma k_F$. RPA calculations were performed using  $\frac{ m^*}{\pi} V_0 = 0.5$. (b) Same as (a) but for the intervalley pairing interaction.
    }
    \label{fig:ideal_R7G_ints}
\end{figure}

As we shall now show, intervalley pairing interactions are suppressed relative to the intravalley pairing interactions. Assuming pairing at a single Fermi surface, we can decompose the pairing interactions into angular momentum sectors, $u^{\tau\tau'}_{\theta}=\sum_n u^{\tau\tau'}_n e^{in\theta}$, where $\theta$ is the angle formed by two points, $\bf k$ and $\bf k'$ on the Fermi surface. The dominant pairing sector corresponds to the most negative $u^{\tau\tau'}_n$, and leads to superconductivity with angular momentum $n$, and a gap function that winds $n$-times. The projected pairing interaction can be decomposed as
\begin{equation}
    u^{\tau \tau'}_n = \sum_{m,m'} \lambda^\tau_{m}\lambda^{\tau'}_{m'} v_{n-m-m'}.\label{eq:PairingIntExpansion}
\end{equation}
where $\lambda^\tau_n$, and $v_n$ are the Fourier transforms of $\Lambda^\tau_{\bf k k'}$ and $V^{\rm{RPA}}_{\mathbf{k}-\bf{k}'}$, with respect to $\theta$, the angle formed by the points $\bf k$ and $\bf k'$ on the Fermi surface.

To proceed, we note that $V^{\rm RPA}_{\bf q}\geq0$ for all momenta, such that $v_0$ is positive and large. This occurs because, in real space, the overscreened Coulomb interaction remains strongly repulsive at short distances.
If quantum geometry plays no role
($\lambda^\tau_m = \delta_{m,0}$), the $v_0$ repulsion only affects the $n = 0$ sector. However, from Eq.~\eqref{eq:PairingIntExpansion} we can see that the $v_0$ repulsion will also suppress $n \neq 0$ sectors when there is non-trivial quantum geometry. We refer to this suppression as \textit{residual repulsion poisoning}.

Let us first consider the intervalley pairing interactions, $u^{+-}_n$.
Comparing Eqs.~\eqref{eq:holomorphicFFs} and~\eqref{eq:PairingIntExpansion} we find that the $|n|\leq \ell_{\mathrm{max}}$ intervalley sectors will be poisoned by $v_0$, inhibiting pairing in these sectors. The high angular momentum sectors, $|n| > \ell_{\mathrm{max}}$, are unpoisoned, but pairing interactions in high angular momentum sectors are naturally weaker\cite{chubukov1993kohn,katznelson1968introduction}. On the other hand, for intravalley channels only the $2\ell_{\mathrm{max}}\geq n\geq 0$ sectors are poisoned, while the $n<0$ sectors are all unpoisoned~\cite{maymann2025pairingmechanismdictatestopology}.
Notably, this includes the $n = -1$ sector,
where the attractive component of $V^{\rm RPA}$ tends to be the strongest due to quantum geometric underscreening~\cite{Shavit_Alicea_PhysRevLett.134.176001,Shizeng_jahin2025gt8h-czf3}.

To make this discussion concrete, let us consider the explicit holomorphic band found in the ideal limit of R$N$G\cite{tan2025ideallimitrhombohedralgraphene}, the active band of which has $\ell_{\mathrm{max}}  = N-1$ and $c_{\ell} = \gamma^{\ell}$, where $\gamma^{-1} \approx .578$nm$^{-1}$ is the momentum scale where the distribution of Berry curvature is peaked. This ideal limit is a reasonable approximation of R$N$G when $\gamma k_F <  1$\cite{desrochers2026energetics}. For simplicity we will assume a quadratic dispersion, $\mathbf{k}^2/2m^*$, and will restrict our attention to R$7$G, which will also be the focus of the next section. In the SM we show that the trends observed here also occur for other layer numbers and related holomorphic bands~\cite{supp}.

In Fig.~\ref{fig:ideal_R7G_ints} we plot $u^{\tau\tau'}_n$ for ideal R$7$G, taking the
$d\rightarrow 0$ limit of gate-screened Coulomb interaction, where it becomes a contact interaction, $V_{\bf q} = V_0 > 0$.
The intravalley pairing strength exceeds the intervalley one for all values of $\gamma k_F$.
In the $K$ valley, the dominant channel is $n = -1$ for $\gamma k_F < 1$. The resulting superconducting state is valley-helical, in the sense that the gap function has negative chiral winding in the $ K$ valley and positive chiral winding in the $K'$ valley. Such a state displays a number of unusual phenomena, which we will discuss later on.

Fig.~\ref{fig:ideal_R7G_ints} can be easily understood in terms of residual repulsion poisoning. As discussed previously, the intravalley $n = -1$ sector is unpoisoned, and we find that it is indeed the strongest pairing sector. Additionally, we observe repulsive pairing interactions in the $ n\geq 0$ intravalley sectors, and in both positive and negative intervalley sectors, as expected. There is very small attraction in the intervalley $n = \pm 5$ sector, $\frac{m^*}{\pi} u^{+-}_{\pm 5} \sim -10^{-4}$, but it is highly suppressed compared to the intravalley $n = -1$ sector.

The behavior observed in Fig.~\ref{fig:ideal_R7G_ints} can be captured analytically by expanding the terms in Eq.~\eqref{eq:PairingIntExpansion} in powers of $\kappa \equiv \gamma k_F$. The leading-order contributions are
\begin{equation}\begin{split}
    u^{+ -}_0 =  v_0, \phantom{=}
    u^{+ -}_{\pm 1} =  \kappa^2 v_0 -  \frac{\kappa^2 \Pi_0 V^2_0}{\left(1+\Pi_0 V_0\right)^2}
\label{eq:interExpansion}\end{split}\end{equation}
while the leading-order intravalley pairing interactions are
\begin{equation}\begin{split}
    &u^{++}_0 =  v_0, \phantom{=}
    u^{++}_{+ 1} =  2\kappa^2 v_0 -  \frac{\kappa^2 \Pi_0 V^2_0}{\left(1+\Pi_0 V_0\right)^2},\\
    &u^{++}_{- 1} = -\frac{\kappa^2 \Pi_0 V^2_0}{\left(1+\Pi_0 V_0\right)^2}.
\label{eq:intraExpansion}\end{split}\end{equation}
Here, $v_0 = \frac{V_0}{1+\Pi_0 V_0} > 0$ is the short-distance repulsion, and $\Pi_0 = \frac{2m^*}{\pi}$ is the $\bf{q}$ $= 0$ polarizability. The $-\frac{\kappa^2 \Pi_0 V^2_0}{\left(1+\Pi_0 V_0\right)^2}$ contribution is from quantum geometric underscreening~\cite{Shavit_Alicea_PhysRevLett.134.176001,Shizeng_jahin2025gt8h-czf3}.
Comparing Eqs.~\eqref{eq:interExpansion} and~\eqref{eq:intraExpansion}, we find that the strongest pairing is in the intravalley $n = -1$ sector, in agreement with the results in Fig~\ref{fig:ideal_R7G_ints}. While we have focused on ideal R$7$G, Eqs.~\eqref{eq:interExpansion} and~\eqref{eq:intraExpansion} are the leading order behavior for \textit{all} ideal R$N$G.

We note that our present discussion relied on the simplifying assumption of rotation symmetry in each valley. In real systems, there is only discrete rotation symmetry, leading to mixing between different angular momentum sectors.  However, somewhat surprisingly, we find negligibly small angular momentum mixing in our realistic treatment of R$N$G, justifying the angular momentum decomposition used here. As we shall discuss, rotation symmetry breaking--specifically trigonal warping--does have a significant effect on intravalley superconductivity, but this is related to changing the particle-particle susceptibility, not the pairing interactions.

\begin{figure*}
    \centering
    \includegraphics[width=1.0\linewidth]{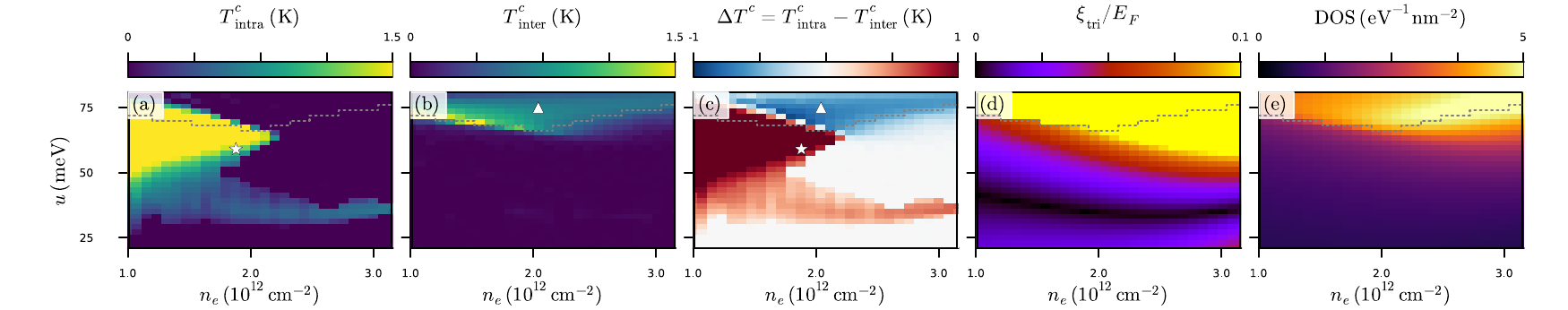}
    \caption{\textbf{SC in time-reversal symmetric R$7$G.} (a) Superconducting transition temperature for intravalley pairing. The dashed line indicates a transition from a normal state with a single Fermi pocket to one with 4 Fermi pockets in each valley (Note: this does not induce a topological transition for the superconducting state) (b) Same as (a) but for intervalley pairing. (c) The difference between the intravalley and intervalley transition temperatures,  $\Delta T^c = T^c_{\mathrm{intra}}-T^c_{\mathrm{inter}}$. (d) The magnitude of trigonal warping at the Fermi level, $\xi_{\rm tri}$, relative to the Fermi-energy, $E_F$. (e) The DOS per spin-valley flavor at the Fermi level. }
    \label{fig:R7G_Tcs}
\end{figure*}

\textit{\textbf {Valley-helical SC in R$N$G.--}}
Having established the possibility of a valley-helical superconducting state in simplified settings, we now turn our attention to realistic models of R$N$G, the Hamiltonian for which is given in the SM~\cite{supp}. Again, we consider a TR-symmetric and spin-degenerate (full metal) normal state. For $4$- through $8$-layers, we find sizable regions of valley-helical SC. Here, we will restrict our attention to R$7$G, and relegate the results on other layer numbers to the SM~\cite{supp}.

In Figs.~\ref{fig:R7G_Tcs}(a-b), we plot the superconducting transition temperatures in the intravalley channel, $T^c_{\mathrm{intra}}$, and intervalley channel, $T^c_{\mathrm{inter}}$, in terms of the total conduction-band filling, $n_e$, and the interlayer potential, $u$, which encodes a displacement field. For the RPA overscreened pairing interaction, we used a bare interaction with $\epsilon_\perp = 6$ and $d = 20$nm. The transition temperatures are obtained by solving the full gap equation in Eq.~\eqref{eq:linearizedGap} for $0 \leq |\bf{k}|$$ \leq 1.5$ nm$^ {-1}$ (the characteristic Fermi momentum is $\lesssim 0.6$ nm$^ {-1}$). Details of the calculations are given in the SM~\cite{supp}.

In Fig.~\ref{fig:R7G_Tcs}(c), we plot  $\Delta T^c =  T^c_{\mathrm{intra}}- T^c_{\mathrm{inter}}$, which shows an extended region of \textit{dominant intravalley pairing}. In almost all of this region, the normal state has one Fermi pocket per valley with low trigonal warping. We quantify the trigonal warping strength by
\begin{equation}
    \xi_{\rm tri}\equiv
    \sum_{ FS } \left|\int_{\mathbf{k}\in \rm FS}\frac{1}{\pi}e^{i 3\theta_{{\bf k}}}\left(
    \xi^{+}_{{\bf k}} - \xi^{{+}}_{{-\bf k}}
    \right)\right|,
\label{eq:TrigWarpEnergy}\end{equation}
where the sum is over all Fermi-surface(s) of the system, $\theta_{{\bf k}}$ is the angle formed by the $\bf k$, and the integral is over all $\bf k$ on a given Fermi-surface~\footnote{Because of the $|...|$ in Eq.~\eqref{eq:TrigWarpEnergy}, $\xi_{\rm tri}$ is the same if we were to instead consider the $\tau = -$ valley}.
We plot $\xi_{\rm tri}$ in Fig.~\ref{fig:R7G_Tcs}(d). Comparing Figs.~\ref{fig:R7G_Tcs}(c) and (d), we find that valley-helical SC primarily occurs in regions with $\xi_{\rm tri}\leq 0.1 E_F$, where $E_f$ is the Fermi-energy. As we show in the SM~\cite{supp}, trigonal warping acts like an angular-dependent Zeeman splitting of the Fermi-surface in the intravalley channel. Consequently, intravalley pairing is more prevalent in regions with low $\xi_{\rm tri}$. This explains the ``elephant's trunk" region of intravalley SC, which extends throughout a region where $\xi_{\rm tri}$ is nearly zero.

For larger $u$, there is a transition from dominant intravalley SC to dominant intervalley SC. This coincides with a transition into a four-Fermi-pocket normal state. The four-Fermi-pocket normal state has much higher trigonal warping, inhibiting intravalley SC, but at the same time has a higher density of states (DOS) [Fig.~\ref{fig:R7G_Tcs}(e)], which benefits intervalley SC.

\begin{figure}
    \centering
    \includegraphics[width=1.0\linewidth]{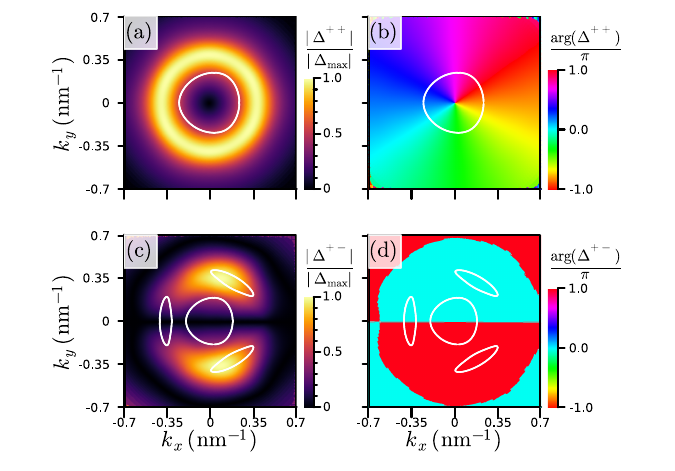}
    \caption{\textbf{R$7$G Intravalley and intervalley gap functions.} (a-b) Amplitude and phase of the leading intravalley gap function in momentum space for $u = 59$meV and $n=1.9\times 10^{12}/$cm$^2$ (star in Fig.~\ref{fig:R7G_Tcs}). The Fermi-surface is shown in white. (c-d) Same as (a-b) but for one of the two degenerate leading intervalley gap functions, at $u = 75$meV and $n=2\times 10^{12}/$cm$^2$ (triangle in Fig.~\ref{fig:R7G_Tcs}). Both the intravalley and intervalley gap functions are spin-triplet.}
    \label{fig:R7G_Gaps}
\end{figure}

Throughout the parameter space in Fig.~\ref{fig:R7G_Tcs}, the leading intravalley gap functions have winding number $-1$ [see Fig.~\ref{fig:R7G_Gaps}(a-b)], matching what we found for ideal R$7$G. The entire red region of Fig.~\ref{fig:R7G_Tcs}(c) therefore realizes valley-helical SC.
As we shall discuss below, the valley-helical SC is also topologically nontrivial. The leading intervalley gap functions come in degenerate pairs due to TR symmetry. Since the intervalley LGE is real, the gaps are real with a nodal structure [see Fig.~\ref{fig:R7G_Gaps}(c-d)].

\textit{\textbf {Phenomenology of valley-helical SC.--}}
We now discuss the unique phenomena of valley-helical SC. Using the conventional Landau framework, the valley-helical phase is defined in terms of two independent order parameters: $\Delta^{++} \equiv \Delta^{K}$ and $\Delta^{--} \equiv \Delta^{K'}$. In addition to carrying charge $2e$, the order parameters also each carry twice the corresponding valley momentum and, due to the chiral pairing, equal and opposite angular momentum.

The angular momentum of the superconducting order parameters couples linearly to an external perpendicular magnetic field, $B_\perp$. In Landau-Ginzburg theory, this leads to a term of the form $B_\perp \left( |\Delta^{K'}|^2-|\Delta^{K}|^2 \right)$. Note that this term is more relevant than the $\propto B_\perp^2$ pair-breaking effect for small fields. Physically, the linear coupling indicates that a small perpendicular magnetic field will \textit{increase} the superconducting transition temperature for one of the two valleys, enhancing the $T^c$ measured in transport.

The intravalley paired state may also be topological. For each valley, one can define a Bogoliubov-de-Gennes (BdG) Chern number, $C_\tau$. While the net Chern number is zero due to TR-symmetry ($C_K = - C_{K'}$), the system can still display $\mathbb{Z}_2$ TR-symmetry-protected topology. This requires $\mathcal{T}^2 = -1$ TR-symmetry, and that $C_\tau$ is odd\cite{schnyder2008classification}. When topological, the valley-helical state has a $\mathcal{T}$-protected pair of counter-propagating helical edge modes~\cite{ZHANG_Kane_Mele_Tritops_PhysRevLett.111.056402, TRITOPS_HAIM20191}. Kramers' theorem guarantees that these modes are gapless as long as TR-symmetry is preserved. Returning to full metal R$7$G, we note that $\mathcal{T}^2 = -1$ due to spin degeneracy. Additionally, since the normal state always has a particle-like Fermi pocket at each valley center, the BdG Chern number is equal to the gap winding number, i.e., $C_K = -1$~\cite{sato2009topological}. The valley-helical SC found in Fig.~\ref{fig:R7G_Tcs} is therefore topological. If the normal state were instead hole-like at the valley center (e.g., an annular Fermi-surface) the valley-helical SC would be topological trivial, although we do not find this to be the case for R$7$G.

Finally, we remark that in the valley-helical state, the composite order parameter $\Delta^{K*}\Delta^{K'}$ is always non-zero. From the symmetry point of view, $\Delta^{K}\Delta^{K'*}$ constitutes a Kekulé bond distortion\cite{chamon2000solitons, bao2021experimental, scheer2023twistronics}; both break the same translation and rotation symmetries, but are invariant under certain combinations of rotations and translations [see SM\cite{supp}].
A translation symmetry-breaking pattern is expected here, since the valley-helical SC is a Larkin-Ovchinnikov (LO)-type pair-density-wave (PDW)~\cite{berg2009striped,  agterberg2020physics}. However, unlike the other LO-type PDWs, only the bond density modulates in the valley-helical SC, and the density profile is constant[see SM~\cite{supp}] ~\cite{sachdev2003colloquium, allais2014density, faye2017interplay}. Valley-helical SC also has composite charge-$4e$ SC order, $\Delta^{4e} = \Delta^{K}\Delta^{K'}$, which is common to LO-type PDWs~\cite{berg2009charge}.

\textit{\textbf {Conclusion.--}}
In this work, we studied superconductivity arising from overscreening of the Coulomb interaction in systems with valleys-contrasting Berry curvatures.
We found that pairing interactions are generically stronger in intravalley channels due to residual repulsion poisoning of the intervalley channels.
This leads to dominant valley-helical SC,
assuming the relevant particle-particle susceptibility is not too dampened, e.g., by trigonal warping effects.
For realistically modeled R$N$G, there are extended parameter regimes where valley-helical SC has the highest transition temperature. The valley-helical SC is also topological in this region, with gapless counter-propagating edge modes.
We note that Stoner spin-valley polarization
may
generally compete with the superconducting states.
Understanding the details of the competition is an important topic for future work.

While we have focused on spin-degenerate valleys, the same mechanism outlined here also lends itself to systems with spin-polarized valleys, such as $K$-valley TMDs~\cite{wu2019topological,devakul2021magic}, and the half-metal phases of rhombohedral graphene. However, for a single spin species, the available DOS for pairing is reduced by a factor of $1/2$.
Nevertheless, we still identify a finite region of half-metal R$7$G with dominant valley-helical SC [see SM~\cite{supp}].

The residual repulsion poisoning discussed here should be generic to purely electronic pairing mechanisms.
Many-body corrections are not expected to significantly change the short-distance behavior of the Coulomb interaction, as electronic screening is ineffective at small scales.
Intravalley superconductivity is therefore expected to be competitive in such pairing scenarios.
However, if the pairing interaction is retarded, most notably if it is phonon-originated, the short-distance repulsion is inconsequential,
and one expects intervalley superconductivity to dominate.

An increase of $T^c$ in the presence of a small but finite perpendicular magnetic field would be positive evidence that a given TR symmetric superconductor has an intravalley character.
Such an increase in $T^c$ has been reported for a nominally TR-symmetric superconductor in hole-doped R$5$G~\cite{LongJu_Magnetic_boosted_SC_Seo2026}, potentially indicating an intravalley pairing.

\begin{acknowledgments}
We thank Daniel Agterberg, Daniel Parker, and Aaron Sharpe for useful conversations.
GS acknowledges support from the Walter Burke Institute for Theoretical Physics at Caltech, and from the Yad Hanadiv Foundation through the Rothschild fellowship. JMM acknowledges support from a Leinweber Institute for Theoretical Physics postdoctoral fellowship at Stanford University. This work was performed in part at the Aspen Center for Physics, which is supported by National Science Foundation grant PHY-2210452.
This material is based upon work supported by the Air Force Office of Scientific Research under award
number FA9550-25-1-0343.
\end{acknowledgments}

\bibliographystyle{apsrev4-2}
\bibliography{refpp}

\clearpage
\onecolumngrid
\begin{center}
\textbf{\large Supplemental Material for ``Valley-Helical Superconductivity Driven by Repulsion and Quantum Geometry: Applications to Time-Reversal Symmetric Rhombohedral Graphene"}
\end{center}

\setcounter{figure}{0}
\renewcommand{\figurename}{Fig.}
\renewcommand{\thefigure}{S\arabic{figure}}
\setcounter{equation}{0}
\renewcommand{\theequation}{S\arabic{equation}}
\setcounter{section}{0}
\renewcommand{\thesection}{S\arabic{section}}
\makeatletter
\@removefromreset{equation}{section}
\makeatother

\section{Details of gauge fixing}
For all the bands analyzed in this work, we used the following gauge for the Bloch wavefunctions. To define the gauge, we first consider the $\bf k$$=0$ Bloch function
\begin{equation}
    \ket{u_0} = \sum_{\ell} c_{\ell,0} \ket{\ell},
\end{equation}
where the $\ket{\ell}$ correspond to different microscopic orbitals that comprise the multi-band system. Let us define $\ell_{0,\text{max}}$ as the $\ell $ with the largest $|c_{\ell,0}|$. We fix the gauge for all $\ket{u_{\bf k}}$ by requiring that $\braket{\ell_{0,\text{max}}}{u_{\bf k}}$ is real and positive. Since $\braket{\ell_{0,\text{max}}}{u_{\bf k}}$ is never zero for all bands considered in this work, this defines a smooth gauge. It can also be directly confirmed that this gauge respects any rotation symmetries of the model, provided that the basis elements, $\ket{\ell}$, transform as irreducible representations of the relevant rotation symmetry group. This is the case for all the basis elements considered here.

\section{Linearized angular gap equation}\label{sec:app_linearized_angular_BCS}
Here, for the sake of completeness, we explain how one obtains the approximate form of the gap equation of the rotationally symmetric gap equation.

The starting point is the full gap equation of the form
\begin{equation}
    \Delta_{\bf k}=
    -\int\frac{d{\bf k'}}{\left(2\pi\right)^2}
    u_{\bf kk'}
    \frac{\tanh{\frac{E_{\bf k'}}{2T}}}{2E_{\bf k'}}
    \Delta_{\bf k'}.
\end{equation}
Notice that positive $u$ here corresponds to a repulsive interaction.

Next, one neglects the radial dependence of $u_{\bf kk'}$, justified in the weak-coupling limit where the key contributions to the gap equation come from the immediate vicinity of the Fermi surface.
Plugging in $E_{\bf k'}=\frac{k'^{2}-k_{F}^{2}}{2m}$,
\begin{equation}
    \Delta_{\theta}=\frac{m}{4\pi}\int_{-E_{F}/2T_{c}}^{E_\Lambda/2T_{c}}dx\frac{\tanh x}{x}\int\frac{d\theta'}{2\pi}u_{{\bf \theta\theta'}}\Delta_{{\bf \theta'}},
\end{equation}
with the momentum cutoff $\Lambda$, and $E_\Lambda=\Lambda^2/\left(2m\right)-E_F$.
By defining an effective energy scale $W\sim\sqrt{E_\Lambda E_{F}}$, we recast the equation as
\begin{align}
    \Delta_{\theta}&=-\frac{m}{2\pi}\log\frac{W}{T^c}\int\frac{d\theta'}{2\pi}u_{\left[\theta,\theta'\right]_{k_F}}\Delta_{\theta'}\nonumber\\
    &=-\log\frac{W}{T}{\cal K}_{\left[\theta,\theta'\right]_{k_F}}\Delta_{\theta'},
\end{align}
with the linear operator
\begin{equation}
    {\cal K}_{\left[\theta,\theta'\right]_{k_F}}\Delta_{\theta'}=\frac{m}{2\pi}\int\frac{d\theta'}{2\pi}u_{\left[\theta,\theta'\right]_{k_F}}\Delta_{\theta'}.\label{eq:Klinearoperatorbcs}
\end{equation}
We write $u_{\left[\theta,\theta'\right]_{k_F}}$ as the interaction between two points on the Fermi surface (with momentum $k_F$) at angles $\theta$ and $\theta'$.
We label the $n$th eigenvalues of $\cal K$ as $-\lambda_n$, with a corresponding superconducting critical temperature
$T^c\approx W \exp\left(-\frac{1}{\lambda_n}\right)$.

\section{Analytical models}
\subsection{Ideal R$N$G}\label{sec:idealRNG}
Here, we review some of the details of the ideal rhombohedral $N$-layer graphene (R$N$G) introduced in Ref.~\cite{tan2025ideallimitrhombohedralgraphene}.
The model is inspired by a simplified description of the low-energy single particle properties of the full R$N$G model.
The model operates on a Hilbert space of sublattice ($A/B$) and layer number $\ell=1,2,...,N$.
Including parabolic dispersion in the model, the Hamiltonian written in the sublattice basis takes the form
\begin{equation}
    H_{\rm ideal}=\begin{pmatrix}0 & {\cal D}^{\dagger}\\
{\cal D} & -\Delta_{B}\mathbb{1}
\end{pmatrix}-\frac{\nabla^{2}}{2m}\begin{pmatrix}\mathbb{1}\\
 & -\mathbb{1}
\end{pmatrix},
\end{equation}
where $\cal D$ is a $N\times N$ matrix with elements
${\cal D}_{\ell,\ell+1}=-t$, diagonal elements ${\cal D}_{\ell,\ell}=-2iv\partial_{\bar z}$ for $\ell\neq N$, ${\cal D}_{N,N}=0$, and all the other elements are zero.
Here, $\partial_{\bar z}=\frac{1}{2}\left(\partial_x-i\partial_y\right)$ is the anti-holomorphic derivative, $t\approx381$ meV represent the interlayer hopping in R$N$G, and $v\approx1.1\,10^6\frac{\rm m}{\rm sec}$ is the velocity of graphene intralayer Dirac fermions.
We can thus define a typical length scale $\gamma=\hbar v/t_{\perp}\approx1.91$ nm.

In the absence of the parabolic dispersion term in $H_{\rm ideal}$, the model hosts two oppositely-sublattice polarized flat-bands, corresponding to (in the sublattice basis) $|A_{\bf k}\rangle=\left(|\varphi_{\bf k}\rangle,0\right)^T$, and $|B_{\bf k}\rangle=\left(0,|\tilde{\varphi}_{\bf k}\rangle\right)^T$.
Here, $|{\varphi}_{\bf k}\rangle$ and $|\tilde{\varphi}_{\bf k}\rangle$ are the zero modes of $\cal D$ and its adjoint, respectively,
${\cal D}|\varphi_{\bf k}\rangle=0$,
${\cal D^\dagger}|\tilde{\varphi}_{\bf k}\rangle=0$.
The term $\propto\Delta_B$ separates this two bands in energy, and we henceforth assume that the bottom $|B_{\bf k\rangle}$ hole band is filled and inert, and project our interacting physics to the parabolically dispersing $|A_{\bf k}\rangle$ band.

We define $|{\mathsf u}_{\bf k}\rangle\equiv e^{-i{\bf k \cdot r}}|\varphi_{\bf k}\rangle$, which singles out the periodic part of the Bloch wavefunction.
The components of this wavefunction are given by
\begin{equation}
    \left\langle\ell|{\mathsf u}_{\bf k}\right\rangle={\cal N}_{\bf k}
    \left[\gamma\left(k_x+ik_y\right)\right]^{\ell-1},
    \,\,\,\,\,\,
    {\cal N}_{\bf k}=\left(\sum_{\ell=1}^N\left|\gamma{\bf k}\right|^{2\ell-2}\right)^{-\frac{1}{2}}.
\end{equation}
The appropriate wavefunction overlaps can then be computed,
\begin{equation}
    \Lambda_{\bf kk'}={\cal N}_{\bf k}{\cal N}_{\bf k'}\sum_{\ell=1}^N \left[\gamma^2\left(k_x-ik_y\right)\left(k'_x+ik'_y\right)\right]^{\ell-1}.
\end{equation}

It is useful to examine the overlaps between points on a Fermi surface of radius $k_F$, separated by an angle $\theta$,
\begin{equation}
    \Lambda_\kappa\left(\theta\right)=\frac
    {\sum^N_{\ell=1}\kappa^{2\ell}e^{i\left(\ell-1\right)\theta}}
    {\sum^N_{\ell=1}\kappa^{2\ell}}.
\end{equation}
where we defined the dimensionless $\kappa\equiv\gamma k_F$.
The angular harmonic decomposition of the Cooper pair form factors are thus
\begin{equation}
    \lambda_m^K=\frac{\kappa^{2\left(m+1\right)}}{\sum^N_{\ell=1}\kappa^{2\ell}},
\end{equation}
for $0\leq m<N$, and zero otherwise.
The form factors in the opposite valley are given by taking $m\to -m$.

\subsection{Multifold band inversion}
We consider the multiband band inversion model introduced in Ref.~\cite{bandinversion_PhysRevX.14.041040}.
Let us consider the Bloch Hamiltonian
\begin{equation}
    H_{\rm BHZ}=\sum_{{\bf k}}\Psi_{{\bf k}}^{\dagger}\left[\
    vk_xS^x
    +vk_yS^y
    +\left(\frac{k^2}{2m}+M\right)S^z
    \right]\Psi_{{\bf k}},
\end{equation}
with $\Psi_{\bf k}$ a spinor of fermionic annihilation operators at momentum $\bf k$ living in the Hilbert space spanned by the spin-$S$ matrices $S^i$.
We will assign each such model with an integer, $N\equiv 2S$.
We make the judicious choice of $M=-mv^2/2$, such that the spectrum of the $j$ band reads
\begin{equation}
    \epsilon^j_{\bf k}=\left(S-j\right)\frac{k^2+Q^2}{2m},
\end{equation}
where we defined $Q=mv$, and $j=0,1,...,N$ is an integer.
The wavefunction of the $j=0$ band, henceforth the active band, will be our main focus.
Its form-factors are given by,
\begin{equation}
    \Lambda_{{\bf k,k'}}^{\pm}=\left(\frac{Q^{2}+{\bf k\cdot k'}\pm i{\bf k\times k'}}{\sqrt{Q^{2}+\left|{\bf k}\right|^{2}}\sqrt{Q^{2}+\left|{\bf k'}\right|^{2}}}\right)^{N}.\label{eq:multifoldapp}
\end{equation}
Notice that $\pm N$ is the Chern number of this band, and $Q$ can be now thought of as the momentum scale which determines the distribution of the quantum metric $g_{\bf k}$ and the Berry curvature $\Omega_{\bf k}$ in momentum space,
\begin{equation}
    g_{{\bf k}}^{\mu\nu}=\frac{NQ^{2}}{\left(Q^{2}+\left|{\bf k}\right|^{2}\right)^{2}}\delta^{\mu\nu}=\frac{1}{2}\left|\Omega_{{\bf k}}\right|\delta^{\mu\nu}.
\end{equation}
From these overlaps we can extract the Cooper pair form factors for a particular Fermi momentum $k_F\equiv\varkappa Q$.
Their decomposition is of the form
\begin{equation}
    \lambda_m^K=\begin{pmatrix}N\\
m
\end{pmatrix}\frac{\varkappa^{2m}}{\left(1+
    \varkappa^2\right)^N},
\end{equation}for $0\leq m\leq N$, and zero otherwise.

\subsection{Infinite Chern band}
As discussed in Ref.~\cite{bandinversion_PhysRevX.14.041040}, the large $N$ limit of Eq.~\eqref{eq:multifoldapp} leads to a model with uniform Berry curvature in momentum space.
Specifically, one sets first $Q=\sqrt{2N/{\cal B}}$, and then takes the limit $N\to\infty$.
This produces the lowest-Landau-level like overlaps,
\begin{equation}
    \Lambda_{{\bf k,k'}}^{\pm}=
    \exp\left[-\frac{\cal B}{4}\left(\left|{\bf k-k'}\right|^2\pm 2i {\bf k\times k'}\right)\right].
\end{equation}
It is then straight forward to find the Cooper form-factors in terms of the dimensionless $\Phi\equiv {\cal B}k_F^2$,
\begin{equation}
    \lambda_m^K=e^{-\frac{\Phi}{2}}
    \frac{\left(\Phi/2\right)^m}{m!},
\end{equation}
for $m\geq0$.
Notice that here the ingredients of the decomposition are not bounded, which is a consequence of the unboundedness of the total Berry curvature within a Fermi surface.

\section{Analytics for the infinite Chern band}

Much analytical progress may be achieved when considering the case of the infinite Chern band.
First, the dimensionless RPA screened interaction (with double gated Coulomb interactions),
\begin{equation}
    \tilde{V}_{{\bf q}}\equiv\frac{m}{2\pi}V_{{\bf q}}^{{\rm RPA}}=\frac{\tanh\left|{\bf q}\right|d}{\left|{\bf q}\right|\ell_{{\rm TF}}+N_{f}\frac{\Pi_{{\bf q}}}{\Pi_{{\bf 0}}}\tanh\left|{\bf q}\right|d}.
\end{equation}
The screening in this model takes a rather simple form,
\begin{equation}
    \frac{\Pi_{{\bf q}}}{\Pi_{{\bf 0}}}=e^{-\frac{1}{2}{\cal B}\left|{\bf q}\right|^{2}}.
\end{equation}
Now let us consider a Fermi surface with radius $k_{F}$, and project the dimensionless interaction onto it, as a function of angular separation $\theta$.
A useful identity is $\left|{\bf q}\right|^{2}=2k_{F}^{2}\left(1-\cos\theta\right)$.
Further defining the Berry flux $\Phi\equiv{\cal B}k_{F}^{2}$, and $\zeta\equiv2k_{F}\ell_{{\rm TF}}$, we obtain
\begin{equation}
    \tilde{V}_{\theta}=\frac{\tanh\left(\sqrt{2\left(1-\cos\theta\right)}k_{F}d\right)}{\sqrt{\frac{1-\cos\theta}{2}}\zeta+N_{f}e^{-\Phi\left(1-\cos\theta\right)}\tanh\left(\sqrt{2\left(1-\cos\theta\right)}k_{F}d\right)}.\label{eq:fulVtheta}
\end{equation}
This expression is rather unwieldy to reconstruct into angular harmonics on its own.
Instead, we will make use of the observation that this RPA-screened interaction is dominated by a repulsive constant component and an attractive p-wave-like component,
\begin{equation}
    \tilde{V}_{\theta}\approx\bar{v}\left(1-\alpha\cos\theta\right).
\end{equation}
By evaluating the full expression, Eq.~\eqref{eq:fulVtheta} at $\theta=0,\pi$, and matching to the approximate relation,
\begin{equation}
    \tilde{V}_{\theta=0}=\frac{1}{\ell_{{\rm TF}}/d+N_{f}}\approx\frac{1}{N_{f}},
\end{equation}
\begin{equation}
    \tilde{V}_{\theta=\pi}=\frac{1}{\zeta+N_{f}e^{-2\Phi}},
\end{equation}
where we assumed $\ell_{{\rm TF}}\ll d$, and $2k_{F}d\gg1$.
After doing the matching, we find
\begin{equation}
    \bar{v}=\frac{1}{2N_{f}}\left(1+\frac{N_{f}}{N_{f}+\zeta e^{2\Phi}}e^{2\Phi}\right),
\end{equation}
\begin{equation}
    \alpha=\frac{2N_{f}\sinh\Phi-\zeta e^{\Phi}}{2N_{f}\cosh\Phi+\zeta e^{\Phi}}.
\end{equation}
Alternative representations of the above expressions,
\begin{equation}
    \bar{v}=\frac{1}{2N_{f}}\left(\frac{N_{f}e^{-\Phi}+\left(N_{f}+\zeta\right)e^{\Phi}}{N_{f}e^{-\Phi}+\zeta e^{\Phi}}\right),
\end{equation}
\begin{equation}
    \alpha=\frac{-N_{f}e^{-\Phi}+\left(N_{f}-\zeta\right)e^{\Phi}}{N_{f}e^{-\Phi}+\left(N_{f}+\zeta\right)e^{\Phi}}.
\end{equation}
We note that their product, which will be useful in a moment, can be written as
\begin{align}
\alpha\bar{v} & =\frac{1}{2N_{f}}\left(\frac{-N_{f}e^{-\Phi}+\left(N_{f}-\zeta\right)e^{\Phi}}{N_{f}e^{-\Phi}+\zeta e^{\Phi}}\right)\nonumber\\
 & \approx\frac{\sinh\Phi}{N_{f}e^{-\Phi}+\zeta e^{\Phi}},
\end{align}
where in the last line we made the approximation $\zeta\ll N_{f}$, which is usually the limit we are interested in.

\begin{figure}
    \centering
    \includegraphics[width=0.5\linewidth]{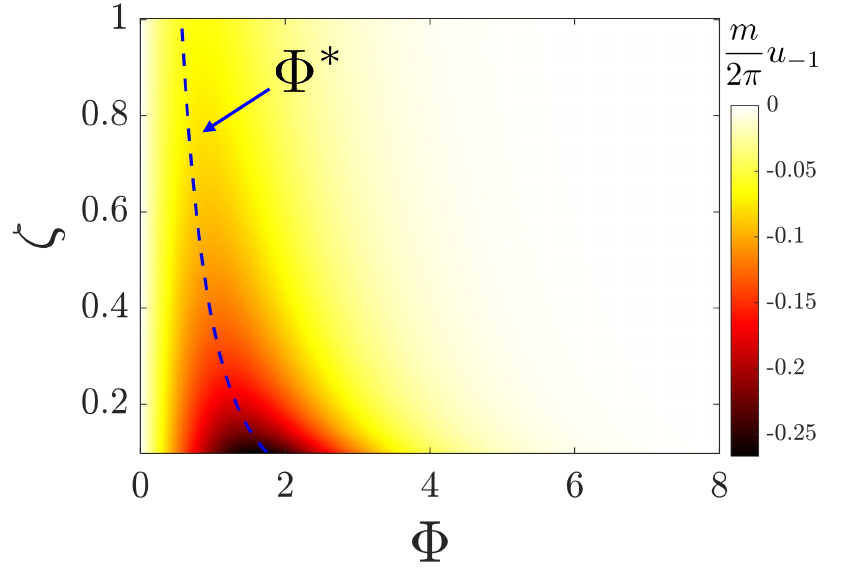}
    \caption{Calculation of the chiral $u_{-1}$, with $N_f=2$.}
    \label{fig:InfchernW}
\end{figure}

Finally, the relevant superconducting coupling in our problem, $u_{-1}$, can be read off from $u_{n}=f_{n}v_{0}+\sum_{m>0}v_{m}\left(f_{n+m}+f_{n-m}\right)$.
The decomposed form factor appropriate for the valley-chiral SC is,
\begin{equation}
    f_{\ell\geq0}=2 e^{-\Phi}\frac{\Phi^{\ell}}{\ell!},
\end{equation}
and vanishes for $\ell<0$.
Thus,
\begin{align}
u_{-1} & =-\alpha\bar{v}f_{0}\nonumber\\
 & =-\frac{\sinh\Phi}{N_{f}+\zeta e^{2\Phi}}.\label{eq:um1analytic}
\end{align}
We can examine the behavior of Eq.~\eqref{eq:um1analytic} in two limits.
Initially, at small $\Phi$, $u_{-1}$ becomes more negative, as the quantum geometric underscreening effect becomes increasingly stronger.
At large enough $\Phi$, the Cooper-pair form factor, contributing an exponential $e^{-\Phi}$ cuts off this behavior, and the instability in this channel becomes very weak.
One may estimate the optimal $\Phi^*$ as the value where the linear in $\Phi$ initial behavior crosses the exponential decay,
and consequently the optimal $u^*_{-1}$,
\begin{equation}
    \Phi^* = {\mathbb W}\left(\frac{N_f}{2\zeta}\right),
\end{equation}
where ${\mathbb W}\left(x\right)$ is the Lambert W function.
As shown in Fig.~\ref{fig:InfchernW}, this approximation faithfully captures the optimal Berry curvature enclosed by the Fermi surface in this model.

Conversely, for the more conventional intervalley superconductor,
\begin{equation}
    u_{n}=2\bar{v}e^{-\Phi}\left[1-\alpha\partial_{\Phi}\right]I_{n}\left(\Phi\right).
\end{equation}
The repulsion is strongly folded in: it is the positive unity in the operator acting on the Bessel function.

\section{Additional numerical results for the analytical models}\label{sec:morenumericalapp}

\begin{figure}
    \centering
    \includegraphics[width=1\linewidth]{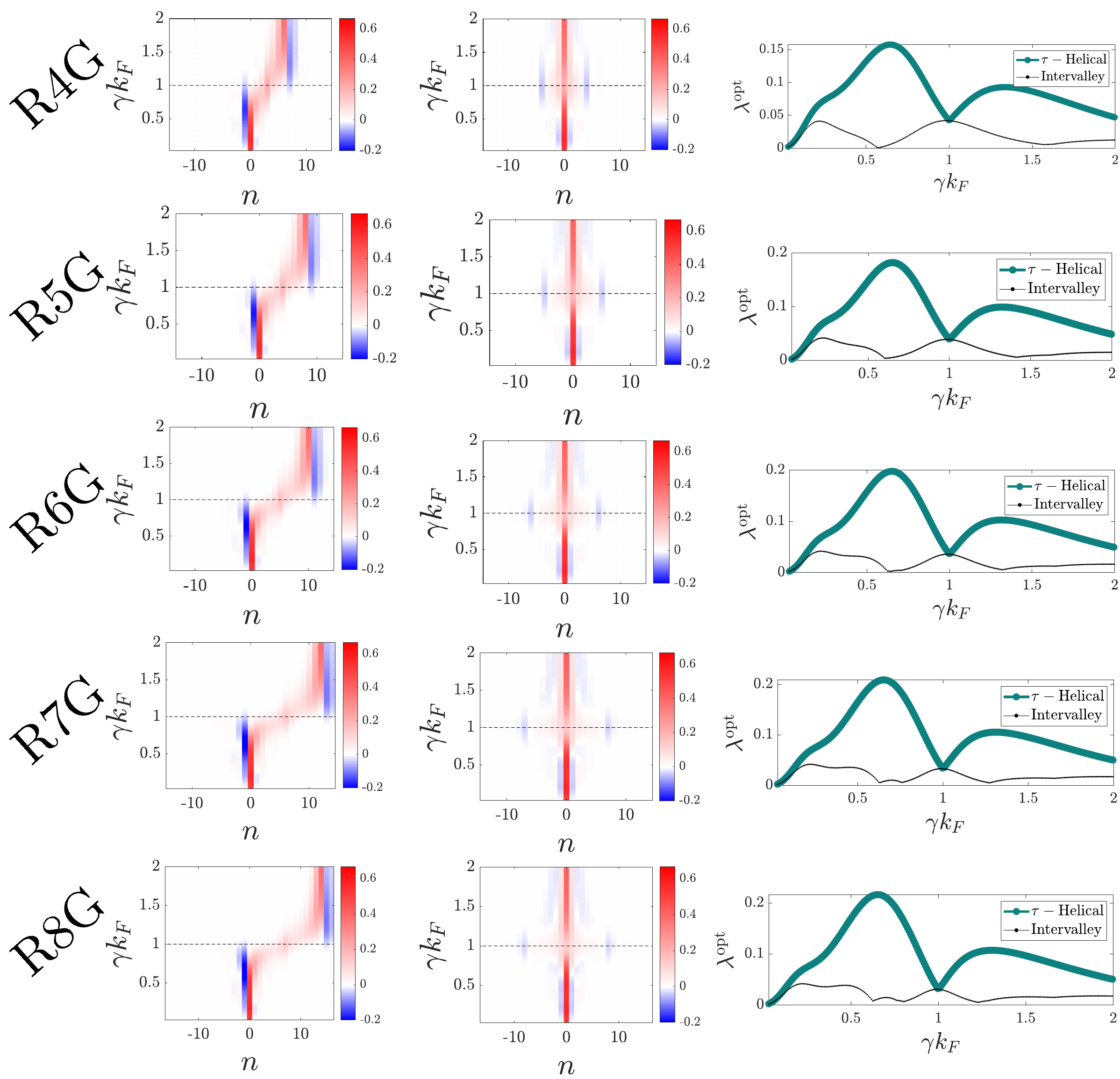}
    \caption{\textbf{Valley-helical SC overwhelms intervalley SC for ideal RNG.}
    The layer number for each row is indicated on the left.
    Right column:
    Optimal (over all relevant $n$ channels) superconducting coupling constant, for the $\tau$-helical SC (green), and for the intervalley superconductor (black).
    The dominant superconducting state is emphasized for each $k_F$.
    Left column:
    Angular decomposition of the various superconducting channels for the $\tau$-helical superconductor.
    We plot $-\lambda_n$ such that pairing instabilities are marked by negative values.
    Middle column:
    Same as the left column, for the intervalley superconductor.
    For all panels we use $\ell_{\rm TF}=0.3\gamma$, $d=8\gamma$.
    }
    \label{fig:numerical_RNGSM}
\end{figure}

\begin{figure}
    \centering
    \includegraphics[width=0.8\linewidth]{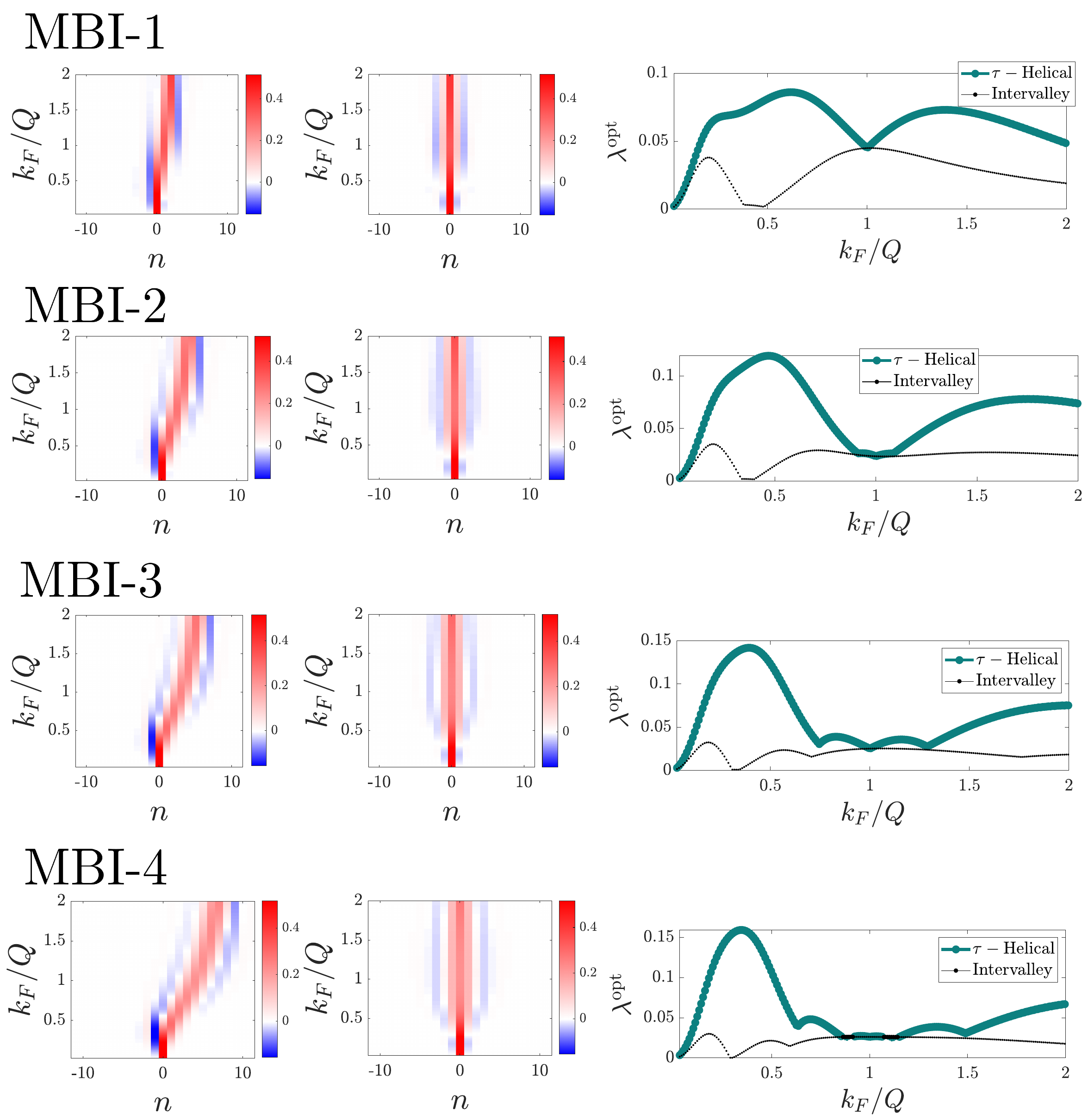}
    \caption{\textbf{Valley-helical SC overwhelms intervalley SC in the multifold band inversion model.}
    Calculations are done using the $N$-fold band inversion model, with $N$ indicated on the left, with varying density (captured by the dimensionless parameter $k_F/Q$).
    Right column:
    Optimal (over all relevant $n$ channels) superconducting coupling constant, for the $\tau$-helical SC (green), and for the intervalley superconductor (black).
    The dominant superconducting state is emphasized for each $k_F$.
    Left column:
    Angular decomposition of the various superconducting channels for the $\tau$-helical superconductor.
    We plot $-\lambda_n$ such that pairing instabilities are marked by negative values.
    Middle column:
    Same as the left column, for the intervalley superconductor.
    For all panels we use $Q\ell_{\rm TF}=0.2$, $Qd=6$.
    }
    \label{fig:numerical_N1234}
\end{figure}

\begin{figure}
    \centering
    \includegraphics[width=0.6\linewidth]{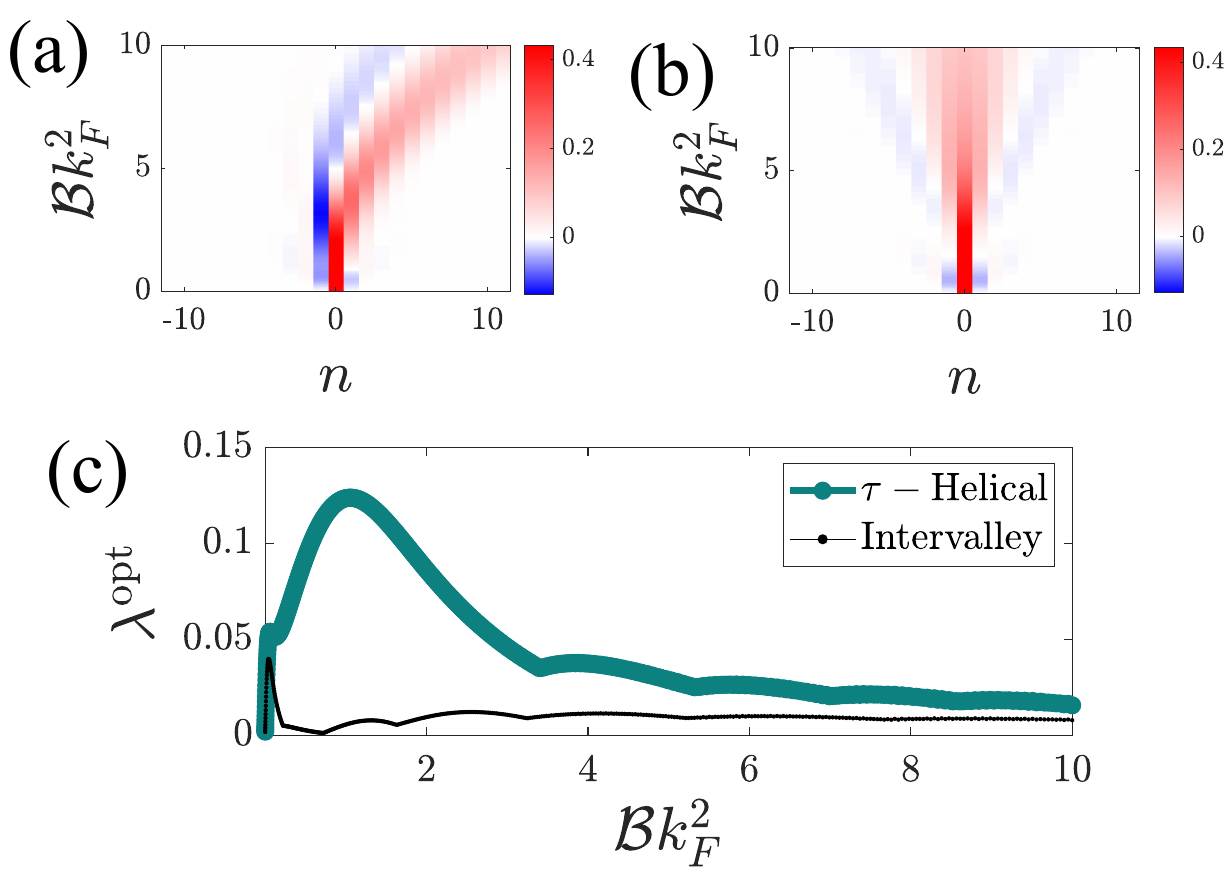}
    \caption{\textbf{Pairing channels in the infinite Chern band model.}
    (a)
    The angular decomposition of $-\lambda_n$ for the valley-helical SC, as a function of the dimensionless $\Phi={\cal B}k_F^2$.
    (b)
    Same as (a), for the intervalley channels.
    (c)
    The optimal coupling constant for both types of SC.
    Here, $d=7\sqrt{\cal B}$, $\ell_{\rm TF}=0.3\sqrt{\cal B}$.}
    \label{fig:numericalNinf}
\end{figure}

\section{Trigonal warping effects}
Let us consider intravalley pairing at a single Fermi-surface with finite $\xi_{\rm tri}$. Trigonal warping will, in general, affect both the pairing interaction and particle-particle susceptibility in Eq.~\eqref{eq:linearizedGap}. However, the latter effect is more important, as it removes the $T\rightarrow 0$ divergence of the particle-particle susceptibility (i.e., the Cooper logarithm) in the intravalley channel. Let us therefore assume trigonal warping only affects the particle-particle susceptibility, where it behaves like an angle-dependent Zeeman splitting.

We begin by assigning an energy scale to the trigonal distortion of an otherwise circular Fermi surface,
\begin{equation}
    \xi_{\rm tri}\equiv
    \left|\int_{\mathbf{k}\in \rm FS}\frac{1}{\pi}e^{i 3\theta_{{\bf k}}}\left(
    \xi^{+}_{{\bf k}} - \xi^{{+}}_{{-\bf k}}
    \right)\right|,
\label{eq:TrigWarpEnergy}\end{equation}
where $\int_{\rm FS}$ represents integration over the Fermi surface, and $\xi{\bf k}$ is the dispersion.
The linearized equation for the superconducting critical temperature in the relevant channel,can thus be approximated as
\begin{equation}
    \frac{1}{\lambda}=\int\frac{d\xi}{2\xi}\int\frac{d\theta}{2\pi}
    \sum_{\sigma=\pm}
    \tanh\left(\frac{\xi+\sigma\xi{{\rm tri}}\cos3\theta}{2T^c}\right),
\end{equation}
where $\lambda$ is the effective superconducting coupling constant.
This equation can be transformed into a transcendental equation for $T^c$, in terms of its relation to the rotationally-symmetric critical temperature $T^{c0}=T^c\left(\xi{\rm tri}=0\right)$,
\begin{equation}
    \log\left(\frac{T_{c}^{0}}{T_{c}}\right)=\int_{0}^{2\pi}\frac{d\phi}{2\pi}\Re\left\{ \psi\left(\frac{1}{2}+i\frac{\xi{{\rm tri}}\cos\phi}{2\pi T_{c}}\right)-\psi\left(\frac{1}{2}\right)\right\}.
\end{equation}
Performing the angular integral explicitly, and defining the dimensionless parameter $\tilde{\epsilon}=\xi{\rm tri}/T^{c0}$, the equation takes the form
\begin{equation}
    \log\left(\frac{T_{c}^{0}}{T_{c}}\right)=\sum_{n=0}^{\infty}\left(\frac{1}{n+\frac{1}{2}}-\frac{1}{\sqrt{\left(n+\frac{1}{2}\right)^{2}+\left(\frac{\tilde{\epsilon}}{2\pi}\frac{T_{c}^{0}}{T_{c}}\right)^{2}}}\right).\label{eq:TCtrigonalSUPP}
\end{equation}
For the sake of comparison, the resulting equation for a conventional pair-breaking Zeeman like perturbation (which does not average out to zero on the Fermi surface),
\begin{equation}
    \log\left(\frac{T_{c}^{0}}{T_{c}}\right)_{{\rm Zeeman}}=\sum_{n=0}^{\infty}\left(\frac{1}{n+\frac{1}{2}}-\frac{n+\frac{1}{2}}{\left(n+\frac{1}{2}\right)^{2}+\left(\frac{\tilde{h}}{2\pi}\frac{T_{c}^{0}}{T_{c}}\right)^{2}}\right),
\end{equation}
where $\tilde{h}=h/T^{c0}$ and $h$ is the pair-breaking field.

The critical value of $\xi{\rm tri}$ may be extracted by taking the $T^c\to 0$ asymptotics of Eq.~\eqref{eq:TCtrigonalSUPP},
\begin{equation}
    \log\left(\frac{T_{c}^{0}}{T_{c}}\right)\approx\log\left(2\frac{\tilde{\epsilon}}{2\pi} \frac{T_{c}^{0}}{T_{c}}\right)+\gamma,
\end{equation}
where $\gamma$ is the Euler–Mascheroni constant.
Thus, the critical value is
\begin{equation}
    \xi{\rm tri}^{\left(c\right)} = \pi e^{-\gamma}T^{c0}\approx1.76T^{c0}.
\end{equation}

\begin{figure}
    \centering
    \includegraphics[width=0.45\linewidth]{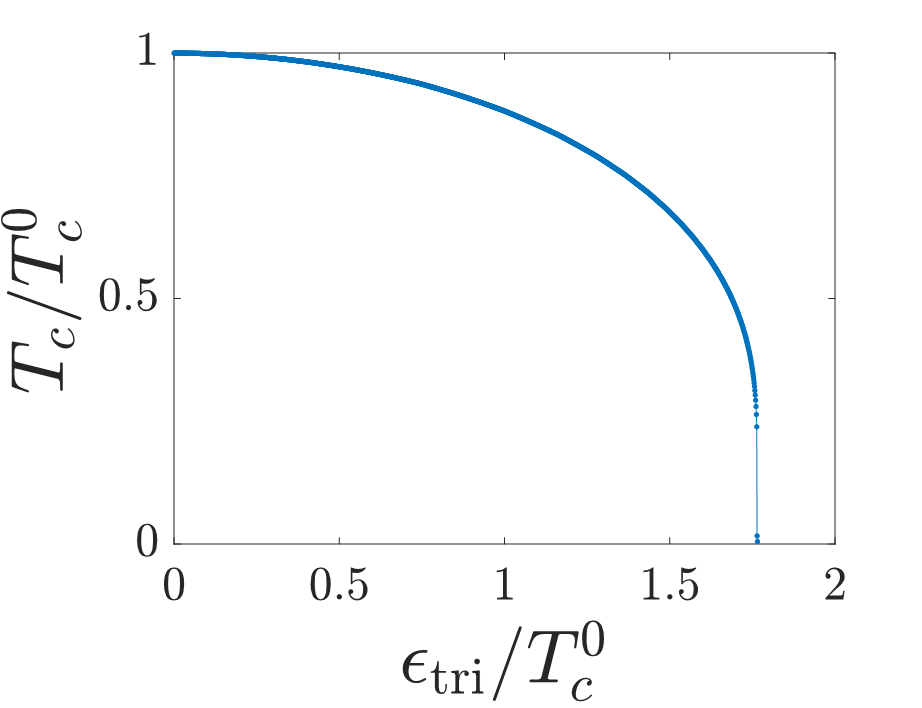}
    \caption{
    Suppression of the superconducting critical temperature as a function of the trigonal distortion energy scale, see Eq.~\eqref{eq:TCtrigonalSUPP}.}
    \label{fig:tridigamm}
\end{figure}

Alternatively, the leading behavior for small trigonal warping, $\tilde{\epsilon}\ll1$, may be extracted by taking this limit for Eq.~\eqref{eq:TCtrigonalSUPP},
\begin{equation}
    \log\left(\frac{T_{c}^{0}}{T_{c}}\right)\approx\frac{1}{2}\left(\frac{\tilde{\epsilon}}{2\pi}\frac{T_{c}^{0}}{T_{c}}\right)^{2}\sum_{n=0}^{\infty}\frac{1}{\left(n+\frac{1}{2}\right)^{3}},
\end{equation}
which in this limit may again be rearranged,
\begin{equation}
    T_{c}\approx T_{c}^{0}\left[1-\left(0.326\frac{\xi{{\rm tri}}}{T_{c}^{0}}\right)^{2}\right].
\end{equation}

\section{Single particle Hamiltonian for $N$-layer rhombohedral graphene}

The real- and reciprocal-lattice vectors of a single graphene layer are
\begin{equation}
\bR_1 = (a,0), \qquad \bR_2 = \left(\tfrac12,\tfrac{\sqrt3}{2}\right)a, \qquad
\bG_1 = \frac{2\pi}{a}\left(1,-\tfrac{1}{\sqrt3}\right), \qquad
\bG_2 = \frac{2\pi}{a}\left(0,\tfrac{2}{\sqrt3}\right),
\end{equation}
with $a\approx\SI{2.46}{\angstrom}$ the graphene lattice constant. The two inequivalent corners of the Brillouin zone are $\bK=\tfrac23\bG_1+\tfrac13\bG_2$ and $\bK'=-\bK$, labeled by a valley index $\tau=K,K'\equiv\pm$.

We consider momentum $\bk=(k_x,k_y)$, measured relative to $\tau\bK$, and define
\begin{equation}
k \equiv k_x + i k_y,
\end{equation}
and $\bar{k} = k^*$.

Consider $N$-layer rhombohedral (ABC) stacked graphene. Indexing the layers as $l=1,\dots,N$ (layer $1$ at the bottom, layer $N$ at the top), the single-spin, valley-$\tau$ Bloch Hamiltonian can be written as an $N\times N$ block matrix
\begin{equation}
h_0^\tau(\bk) =
\begin{pmatrix}
\mathcal  D^\tau_1(\bk) & \mathcal V^\tau(\bk) & \mathcal W & 0 & \cdots & 0 \\
\mathcal V^{\tau\dagger}(\bk) & \mathcal  D^\tau_2(\bk) & \mathcal V^\tau(\bk) & \mathcal W & \ddots & \vdots \\
\mathcal W^\dagger & \mathcal V^{\tau\dagger}(\bk) & \mathcal  D^\tau_3(\bk) & \mathcal V^\tau(\bk) & \ddots & 0 \\
0 & \mathcal W^\dagger & \mathcal V^{\tau\dagger}(\bk) & \ddots & \ddots & \mathcal W \\
\vdots & \ddots & \ddots & \ddots & \mathcal  D^\tau_{N-1}(\bk) & \mathcal V^\tau(\bk) \\
0 & \cdots & 0 & \mathcal W^\dagger & \mathcal V^{\tau\dagger}(\bk) & \mathcal  D^\tau_N(\bk)
\end{pmatrix}.
\label{eq:blockH}
\end{equation}
where each block corresponds to the $A,B$ sublattices of a given layer. The Hamiltonian is parameterized by 7 coupling constants, $\gamma_{0}$, $\gamma_{1}$, $\gamma_{2}$, $\gamma_{3}$, $\gamma_{4}$, $\delta$ and $\Delta_2$

Each layer hosts a monolayer-graphene-like Dirac cone plus an onsite potential. Restricting our attention to the $\tau = +$ valley
\begin{equation}
\mathcal D^+_l(\bk) = \hbar v_0 \begin{pmatrix} 0 &   \bar{k} \\   k & 0 \end{pmatrix} + \mathcal U_l,
\qquad
\hbar v_0 \equiv \frac{\sqrt3}{2}a\,\gamma_0,
\end{equation}
where $\gamma_0$ is the intralayer nearest-neighbor hopping.
The nearest-layer hopping terms are given by
\begin{equation}
\mathcal V^+(\bk) = \begin{pmatrix} -\hbar v_4\,  \bar{k} & -\hbar v_3\,  k \\ \gamma_1 & -\hbar v_4\,  \bar{k} \end{pmatrix},
\qquad
\hbar v_3 \equiv \frac{\sqrt3}{2}a\,\gamma_3, \qquad
\hbar v_4 \equiv \frac{\sqrt3}{2}a\,\gamma_4.
\end{equation}
The next-nearest-layer hopping terms are given by
\begin{equation}
\mathcal W = \begin{pmatrix} 0 & \gamma_2/2 \\ 0 & 0 \end{pmatrix},
\end{equation}
which couples $A_l$ to $B_{l+2}$.

The onsite potential term takes the form, $U_l = \mathrm{diag}\big[u_l^A, u_l^B\big]$, where
\begin{equation}
u_l^A = \begin{cases} \Delta_2 + V_l, & l=1 \\ -\Delta_2+\delta+V_l, & 1<l<N \\ \Delta_2+\delta+V_l, & l=N \end{cases}
\qquad
u_l^B = \begin{cases} \Delta_2+\delta+V_l, & l=1 \\ -\Delta_2+\delta+V_l, & 1<l<N \\ \Delta_2+V_l, & l=N. \end{cases}
\label{eq:onsite}
\end{equation}
where $V_l$ is the layer-resolved potential from the external displacement field. We use $V_{l}= u (N + 1 - 2l) / (N - 1)$, and use $u$ as a control knob.

The $\tau = -$ valley is related to the $\tau = +$ valley by
\begin{equation}
h_0^{-}(\bk) = h_0^{+}(k_x,-k_y).
\end{equation}

Calculations were performed using the parameters in Table~\ref{tab:modelparams}.

\begin{table}[h]
\centering
\begin{tabular}{ccccccc}
\toprule
 $\gamma_0$ (meV) & $\gamma_1$ (meV) & $\gamma_2$ (meV)  & $\gamma_3$ (meV) & $\gamma_4$ (meV) & $\delta$ (meV) &  $\Delta_2$ (meV)\\
\midrule
 $3100$ & $380$ & $-15$  & $290$ & $141$ & $10.5$ & $2$ \\
\bottomrule
\end{tabular}
\caption{Tight-binding parameters for the $N$-layer continuum model.}
\label{tab:modelparams}\end{table}

\section{Details of numeric solutions to the linearized gap equation}

Numeric calculations of the linearized gap equation for R$N$G were performed using $49211$ momentum points sampled over $0 \leq |\bf{k}|$$ \leq 1.5$ nm$^ {-1}$. Momentum points near the bottom the R$N$G bands, $|\bf k| \lesssim 0.6$ nm$^ {-1}$) were sampled more densely. The leading eigenvalues and eigenvectors were found using the Lanczos algorithm.

For the RPA overscreened interaction, we calculated the polarization $\Pi_{\bf q}$ using the expression
\begin{equation}
    \Pi_{\bf q}= - \sum_{\sigma \tau,{\bf k}}    \left|\Lambda^\tau_{\bf k,k+q}\right|^2
   \frac{ f(\xi^\tau_{\bf k+q}) - f(\xi^\tau_{\bf k}) }    {\xi^\tau_{\bf k+q} - \xi^\tau_{\bf k}}.
\end{equation}
For the gap equation of full metal R$N$G, we calculated $\Pi_{\bf q}$ at zero temperature for $ 20,701$ $\bf q$ points equally distributed over $0 \leq |\bf{q}|$$\leq 2.0$ nm$^{-1}$. For each value of $\bf q$, we performed the $\bf k$ summation numerically using $81,165$ momentum points equally distributed in $0 \leq |\bf k$$- \bf{q}|$$\leq 0.8$ nm$^{-1}$. Increasing the integration range beyond this did not change the results. Using these $\bf q$ points, we constructed a continuous polarization function using linear interpolation. The interpolated function was used for the gap equation calculations. For half-metal R$7$G, we used $ 126,937$ $\bf q$ points and $283,991$ $\bf k$ points to account for the smaller pairing scale of the half-metal. We checked that using more  $\bf q$  and $\bf k$ did not meaningfully change the results for full metal R$N$G.

\section{Angular decomposition of microscopically modeled R$7$G}
Here, we quantify the effects of rotation symmetry breaking on gap functions of microscopically modeled R$7$G. With continuous rotation symmetry, the gap functions have a well-defined angular momentum $n$. With trigonal warping and other rotation symmetry-breaking effects, different angular momentum sectors start to mix with each other. To quantify this, we first quantify the winding of the gap function in terms of
\begin{equation}
    \Delta_n = |\frac{1}{A}\sum_{\bf{k}} \Delta(\mathbf{k}) e^{-i \theta_{\bf{k}}}|
\label{eq:angleDecomp}\end{equation}
where $\Delta(\mathbf{k})$ is a given solution to the linearized gap equation, normalized to have a maximum of $1$, and $\theta_{\bf{k}}$ is the angle made by $\bf{k}$.

We define the maximal component for intravalley, $n_{\mathrm{intra,max}}$, as the $n$ with the maximal $\Delta_n$. For intervalley, the gap equation is real, so that the solutions to the eigenvalue problem are also real. Therefore intervalley gap functions, $\Delta_n = \Delta_{-n}$, and we only need to consider $n\geq 0$.

To quantify the effects of band mixing, we introduce the following weights,
\begin{equation}
    w[n_{\mathrm{intra,max}}] = \frac{|\Delta_{n_{\mathrm{intra,max}}}|^2}{\sum_{n} |\Delta_{n}|^2}
\label{eq:intraWeight}\end{equation}
and
\begin{equation}
    w[n_{\mathrm{intra,max}}] = \frac{2|\Delta_{n_{\mathrm{inter,max}}}|^2}{\sum_{n} |\Delta_{n}|^2},
\label{eq:interWeight}\end{equation}
where the extra factor of $2$ accounts for the degeneracy between $\Delta_{n_{\mathrm{inter,max}}}$ and $\Delta_{-n_{\mathrm{inter,max}}}$. Both of the weights are equal to $1$ for a rotationally invariant system. Strong deviations from $1$ are a sign of angular momentum mixing.

\begin{figure*}
    \centering
    \includegraphics[width=1.0\linewidth]{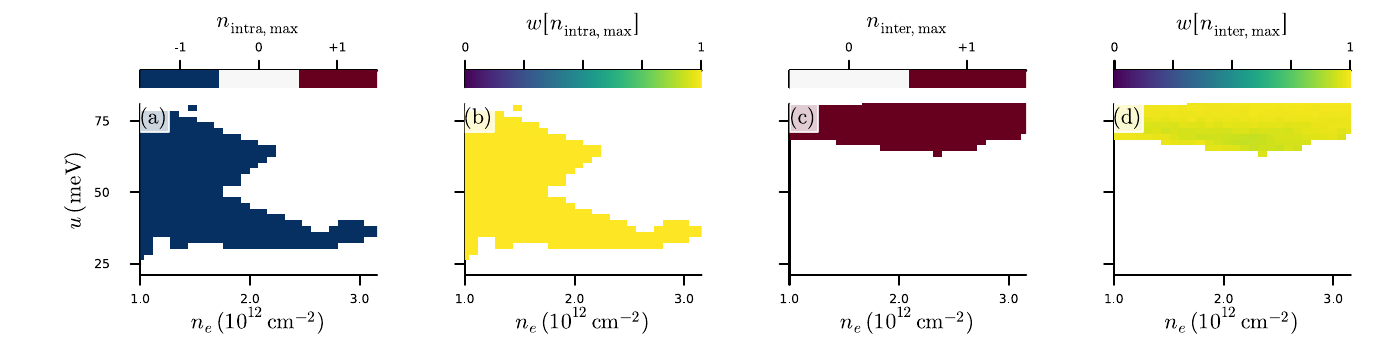}
    \caption{\textbf{Angular decomposition of the R$7$G gap functions.} (a) The maximal component of the angular decomposition in Eq.~\ref{eq:angleDecomp} for the $T = 0 $ intravalley gap functions. Only parameters with a $T^c_{\text{intra}}>.1$K are shown.  (b) the weight of the maximal component for intravalley pairing, defined as in Eq.~\ref{eq:intraWeight} (c) same as (a) but for intervalley gap functions (d) same as (b) but using Eq.~\ref{eq:interWeight} }
\label{eq:R7G_Angle_Decomp}\end{figure*}

In Fig.~\ref{eq:R7G_Angle_Decomp} we plot the maximal components and their respective weight. As we can see, the weights are always close to $1$, indicating that angular momentum mixing is very weak for the gap functions.

\section{Results for other $4-$, $5-$, $6-$, and $8-$layer rhombohedral graphene}
Here we present results for intravalley and intervalley superconductivity in microscopically modelled R$N$G for $N$= $4$, $5$, $6$ and $8$. All calculations follow the same procedure used for the R$7$G results presented in the main text.

\begin{figure*}
    \centering
    \includegraphics[width=1.0\linewidth]{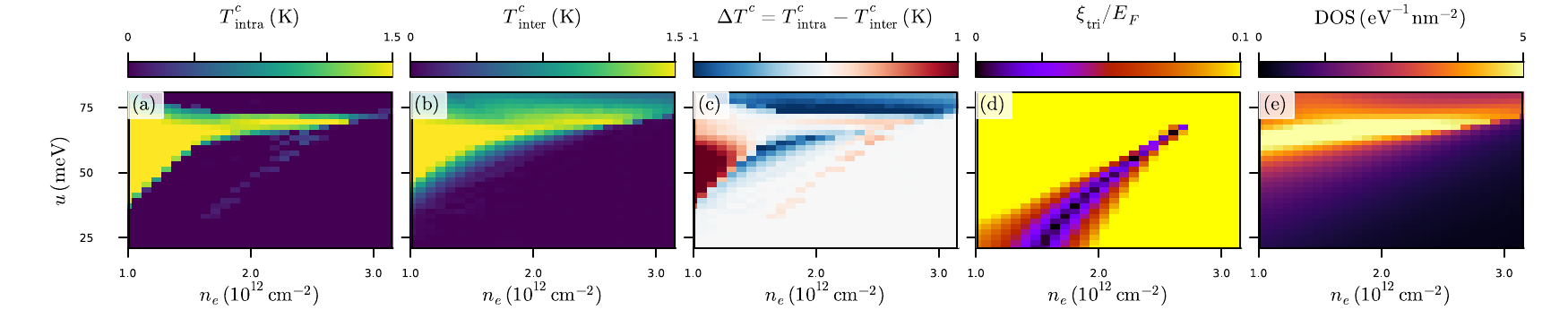}
    \caption{\textbf{SC in time-reversal symmetric R$4$G.} (a) Superconducting transition temperature for intravalley pairing. (b) Same as (a) but for intervalley pairing. (c) The difference between the intravalley and intervalley transition temperatures,  $\Delta T^c = T^c_{\text{intra}}-T^c_{\text{inter}}$. (d) The magnitude of trigonal warping at the Fermi level, relative to the Fermi-energy. (e) The density of states (DOS) per spin-valley flavor at the Fermi level.}
\end{figure*}

\begin{figure*}
    \centering
    \includegraphics[width=1.0\linewidth]{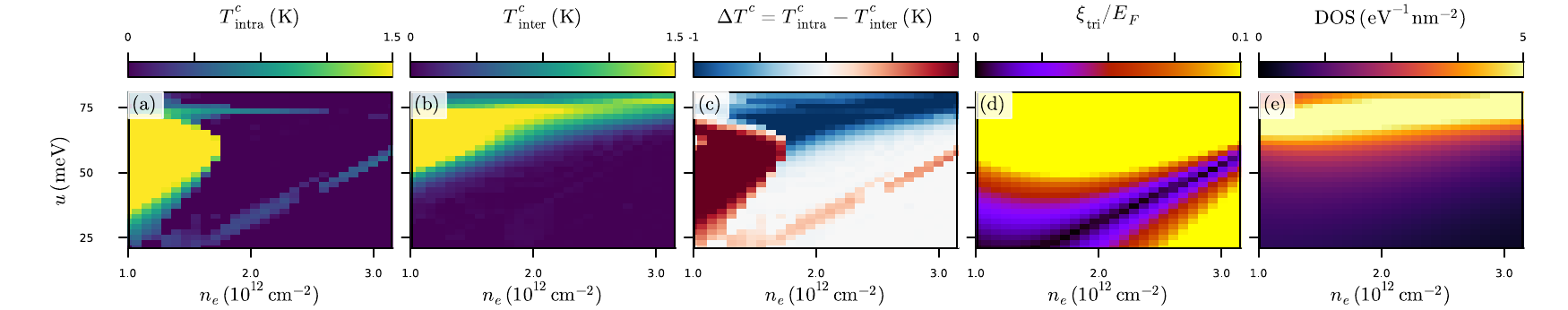}
    \caption{\textbf{SC in time-reversal symmetric R$5$G.} (a) Superconducting transition temperature for intravalley pairing. (b) Same as (a) but for intervalley pairing. (c) The difference between the intravalley and intervalley transition temperatures,  $\Delta T^c = T^c_{\text{intra}}-T^c_{\text{inter}}$. (d) The magnitude of trigonal warping at the Fermi level, relative to the Fermi-energy. (e) The DOS per spin-valley flavor at the Fermi level.}
\end{figure*}

\begin{figure*}
    \centering
    \includegraphics[width=1.0\linewidth]{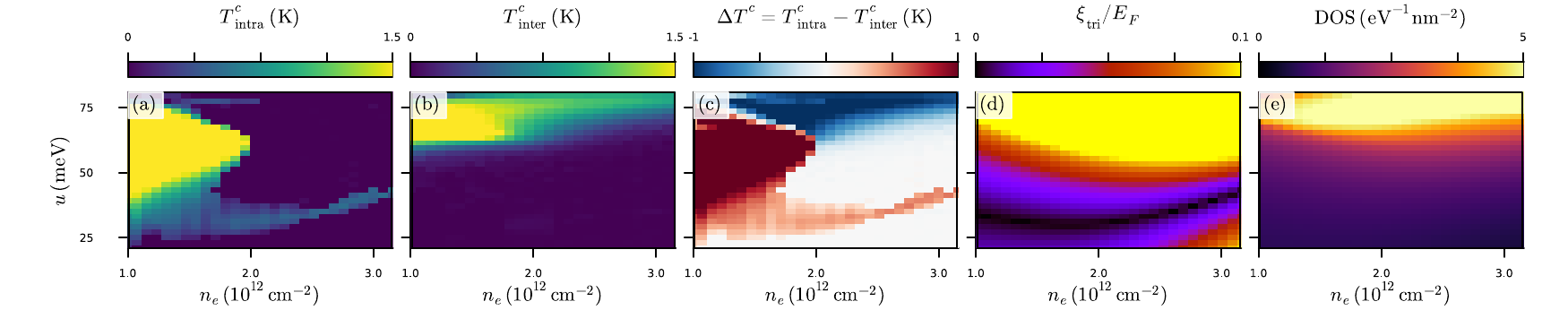}
    \caption{\textbf{SC in time-reversal symmetric R$6$G.} (a) Superconducting transition temperature for intravalley pairing. (b) Same as (a) but for intervalley pairing. (c) The difference between the intravalley and intervalley transition temperatures,  $\Delta T^c = T^c_{\text{intra}}-T^c_{\text{inter}}$. (d) The magnitude of trigonal warping at the Fermi level, relative to the Fermi-energy. (e) The DOS per spin-valley flavor at the Fermi level.}
\end{figure*}

\begin{figure*}
    \centering
    \includegraphics[width=1.0\linewidth]{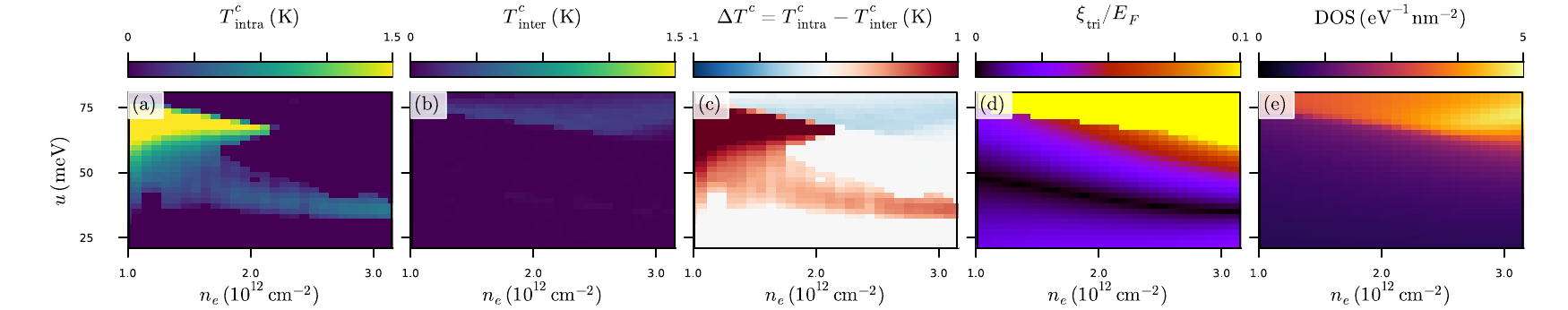}
    \caption{\textbf{SC in time-reversal symmetric R$8$G.} (a) Superconducting transition temperature for intravalley pairing. (b) Same as (a) but for intervalley pairing. (c) The difference between the intravalley and intervalley transition temperatures,  $\Delta T^c = T^c_{\text{intra}}-T^c_{\text{inter}}$. (d) The magnitude of trigonal warping at the Fermi level, relative to the Fermi-energy. (e) The DOS per spin-valley flavor at the Fermi level.}
\end{figure*}

\section{Kekulé-like bond order in valley-helical superconductors}
Here, we will discuss why the valley-helical superconductor displays Kekulé-like bond order, and why it does not induce density modulations, unlike conventional pair-density-waves.

Recall the two superconducting order parameters, $\Delta^K$ and $\Delta^{K'}$, that describe the valley-helical state. These states carry equal and opposite momentum and angular momentum. Under a translation by $ \delta\bm{r}$ (assumed to live on the lattice for graphene), they transform as
\begin{equation}
    \Delta^K \rightarrow \Delta^K e^{i 2 \bm{K}\cdot  \delta\bm{r}}, \phantom{=} \Delta^{K'} \rightarrow \Delta^{K'} e^{-i 2 \bm{K}\cdot  \delta\bm{r}}
\end{equation}
where $\bm{K}$ is the valley momentum. Similarly, for a rotation by $\theta$ that leaves $\bm{K}$ invariant (modulo a reciprocal lattice vector), they transform as
\begin{equation}
    \Delta^K \rightarrow \Delta^K e^{i L\theta }, \phantom{=} \Delta^{K'} \rightarrow \Delta^{K'} e^{-i L\theta }
\end{equation}
where $L$ is the angular momentum of superconductivity in the $K$ valley. For the models considered here $L = -1$, and $\theta$ must be a multiple of $2\pi/3$ for graphene.

Based on the above, consider a composite operation composed of a rotation by $\theta$ followed by translation by $ \delta\bm{r}$. The superconducting order parameters transform as
\begin{equation}
    \Delta^K \rightarrow \Delta^K e^{i L \theta + i 2 \bm{K}\cdot \bm{r}}, \phantom{=} \Delta^{K'} \rightarrow \Delta^{K'} e^{-i L \theta-i 2 \bm{K}\cdot \bm{r}}.
\end{equation}
We can see that this will leave both order parameters invariant, provided that
\begin{equation}
    L \theta  = -2 \bm{K}\cdot  \delta\bm{r} \mod(2\pi)
\label{eq:CompsiteSymmetry}\end{equation}
For graphene, $\bm{K}\cdot  \delta\bm{r}$ is always a multiple of $2\pi /3$, and so for any $ \delta\bm{r}$ it is always possible to find a value of $\theta$ that satisfies Eq.~\ref{eq:CompsiteSymmetry}. This establishes the existence of a set of composite rotation-translation symmetries.

Based on the above, we can identify the composite order parameter $\Delta^{K*}\Delta^K$ with  Kekulé bond order in graphene\cite{gamayun2018valley}. Both order parameters carry momentum $K$ and break rotation symmetry (here defined relative to a site of graphene, not a plaquette), but leave the combination determined by Eq.~\ref{eq:CompsiteSymmetry} invariant.

Importantly, the Kekulé-like bond order is only a periodic distortion of bonds. As we shall now show, the valley-helical state does not support charge fluctuation and has a constant density profile. We will show this by contradiction. Assume there exists charge density wave order in the valley-helical state
\begin{equation}
    \rho_{\bm{Q}} \propto \langle  e^{i\bm{Q}\cdot \bm{r}} \sum_n c^\dagger_n(\bm{r}) c_n(\bm{r}) \rangle
\label{eq:CDW_Definition}\end{equation}
where $c_n(\bm{r})$ is the real-space microscopic annihilation operator for an electron in the $n^\text{th}$ microscopic orbital. Due to the composite symmetry above,
\begin{equation}
    \rho_{R_{\theta}\bm{Q}}  = \rho_{\bm{Q}} e^{i \bm{Q}\cdot \delta\bm{r}}.
\label{eq:densityWaveTrans}\end{equation}
for any $\theta$ and $\delta\bm{r}$ pair that satisfies Eq.~\ref{eq:CompsiteSymmetry}. This can only possibly be true if $\bm{Q} = 2\bm{K}$. However, $\bm{K}$ is invariant under the $\theta$ rotations we are considering here. Therefore,
\begin{equation}
    \rho_{\bm{Q}}  = \rho_{\bm{Q}} e^{i \bm{Q}\cdot \delta\bm{r}},
\end{equation}
so that $\rho_{\bm{Q}} = 0 $. Note that this logic does not preclude bond density waves, as Eq.~\ref{eq:densityWaveTrans} only follows for charge-density-waves, defined as in Eq.~\ref{eq:CDW_Definition}.

\section{Results for half-metal R$7$G}\label{app:halfMetalR7G}
\begin{figure*}
    \centering
    \includegraphics[width=1.0\linewidth]{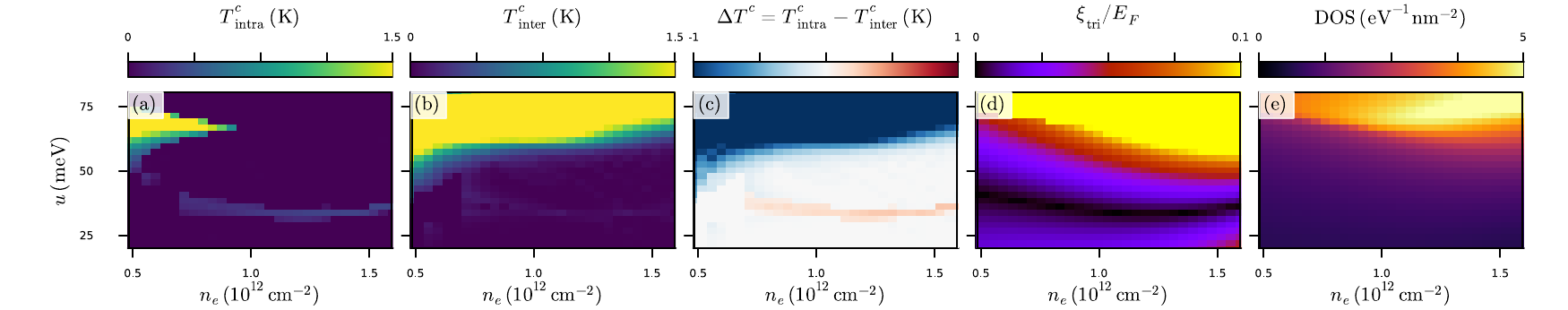}
    \caption{\textbf{SC in R$7$G with spin-polarized valleys.} (a) Superconducting transition temperature for intravalley pairing. (b) Same as (a) but for intervalley pairing. (c) The difference between the intravalley and intervalley transition temperatures,  $\Delta T^c = T^c_{\mathrm{intra}}-T^c_{\mathrm{inter}}$. (d) The magnitude of trigonal warping at the Fermi level, $\xi_{\rm tri}$, relative to the Fermi-energy, $E_F$. (e) The DOS per spin-valley flavor at the Fermi level. }
    \label{fig:R7G_Tcs_half}
\end{figure*}

Here we provide calculations of superconductivity in the half-metal phase of R$7$G where each valley has only a single spin degree of freedom.

The results of the calculations are shown in Fig.~\ref{fig:R7G_Tcs_half} using $\epsilon_\perp = 4$ and $d = 20$nm. We find a strip of valley-helical SC at small $u$. This lies exactly along the wedge of nearly zero trigonal warping, remarked upon in the main text.

\end{document}